\documentclass[12pt]{article}
\usepackage{amsmath,amsthm}
\usepackage{amssymb}
\usepackage{amsfonts}
\usepackage{graphicx}
\usepackage[a4paper,text={150mm,234mm},centering,headsep=8mm,footskip=15mm]{geometry}
\usepackage{hyperref}
\usepackage[font=small,labelfont=bf]{caption}
\usepackage[square,sort,comma,numbers]{natbib}
\usepackage{enumerate}

\newcommand{\beq}{\begin{equation}}
\newcommand{\eeq}{\end{equation}}
\newcommand{\bea}{\begin{eqnarray}}
\newcommand{\eea}{\end{eqnarray}}

\def\d{\mathrm{d}}

\begin{document}

\renewcommand{\theequation}{\thesection.\arabic{equation}} 
\numberwithin{equation}{section} 

\begin{titlepage}
\begin{center}
\rightline{\small ZMP-HH/15-11}
\rightline{\small DESY-15-076}
\vskip 1cm

{\Large \bf Higher-Derivative Supergravity and 
Moduli Stabilization}
\vskip 1.2cm

{\bf  David Ciupke,$^{a}$ Jan Louis,$^{b,c}$ and Alexander Westphal$^{a}$}

\vskip 0.8cm

$^{a}${\em Deutsches Elektronen-Synchrotron DESY, Theory Group, D-22603 Hamburg, Germany}
\vskip 0.3cm

$^{b}${\em Fachbereich Physik der Universit\"at Hamburg, Luruper Chaussee 149, 22761 Hamburg, Germany}
\vskip 0.3cm

{}$^{c}${\em Zentrum f\"ur Mathematische Physik,
Universit\"at Hamburg,\\
Bundesstrasse 55, D-20146 Hamburg, Germany}
\vskip 0.3cm

{\tt david.ciupke@desy.de, jan.louis@desy.de, alexander.westphal@desy.de}

\end{center}

\vskip 1cm

\begin{center} {\bf ABSTRACT } \end{center}

\noindent
We review the ghost-free four-derivative terms for chiral superfields in 
$\mathcal{N}=1$ supersymmetry and supergravity. 
These terms induce cubic polynomial equations of motion for the chiral 
auxiliary fields and correct the scalar potential. 
We discuss the different solutions and argue
that only one of them is consistent with the principles of effective field theory.
Special attention is paid to the corrections 
along flat directions which can be stabilized or destabilized
by the higher-derivative terms. We then compute these 
higher-derivative terms explicitly for the type IIB string 
compactified on a Calabi-Yau orientifold with fluxes via Kaluza-Klein reducing the $(\alpha')^3 R^4$ 
corrections in ten dimensions for the respective $\mathcal{N}=1$ 
K\"{a}hler moduli sector. 
We prove that together with flux and the known $(\alpha')^3$-corrections
the higher-derivative term
stabilizes all  Calabi-Yau manifolds with
positive Euler number, provided the sign of the new correction is
negative.

\vfill
\bigskip

\noindent May, 2015

\end{titlepage}

\newpage

\tableofcontents
\newpage

\section{Introduction}
In many applications supersymmetric field theories or supergravities are
considered as an effective description of a more fundamental theory,
such as string theory. Most properties of this low energy effective theory
are captured by the leading two-derivative Lagrangian $\mathcal{L}_{(0)}$. 
It can, however, happen that specific couplings vanish
in $\mathcal{L}_{(0)}$ and then higher order corrections do become important.
A particular class of corrections are higher-derivative terms
 which
in supersymmetric theories can simultaneously induce
corrections of the scalar potential. 
It is the purpose of this paper 
to analyse supersymmetric higher-derivative operators with this
property
-- both conceptually 
and as a new tool to stabilize moduli in string theory.
Such terms
were also studied in \cite{Cecotti:1986jy, Khoury:2010gb, Koehn:2012ar, Farakos:2012qu, Farakos:2013zsa}, while \cite{Sasaki:2012ka, Koehn:2012np, Farakos:2013cqa, Koehn:2013upa, Gwyn:2014wna} started looking at their implications for cosmology. 

More precisely, we focus on $\mathcal{N}=1$ supersymmetry and supergravity in 
four space-time dimensions and within such theories on ghost-free
higher-derivative operators. 
In non-super\-symmetric theories it is well-known that
the unique ghost-free four-derivative operator for a scalar field $\phi$ is 
given by $(\partial_{\mu} \phi \,\partial^{\mu} \phi)^2$.\footnote{The ghost-free higher-derivative operators are in general those that
do not induce more than two derivatives acting on fields in the equations of motion. Additionally, in supersymmetric theories ghostlike degrees 
of freedom can occur when  a kinetic term for the auxiliary field
is induced \cite{Antoniadis:2007xc, Dudas:2015vka}.} 
Several distinct superspace-operators exist which induce such terms. However, 
there is a unique ghost-free operator given by \cite{Khoury:2010gb}
\begin{equation}
\label{eq:higher_derivative_global_susy}
 \mathcal{L}_{(1)} \sim  
  \int \d^4 \theta \, (D^{\alpha} \Phi)( D_{\alpha} \Phi )(\bar{D}_{\dot{\alpha}} \Phi^{\dagger})( \bar{D}^{\dot{\alpha}} \Phi^{\dagger}) \ ,
\end{equation}
where $D_{\alpha},\bar{D}_{\dot{\alpha}} $ denote the superspace
derivatives, $\d^4 \theta = \d^2 \theta \d^2 \bar{\theta}$ denotes the
integration over the Grassmann variables and $\Phi$ is a chiral 
superfield.
We will see that the equation of motion for the auxiliary field $F$ 
is cubic instead of linear after including $\mathcal{L}_{(1)}$.
This in turn implies up to three inequivalent solutions for $F$ 
and, hence, three inequivalent on-shell theories. The presence 
of this multiplet of theories is 
somewhat puzzling as one seems to loose predictability. 
However, studying the explicit solutions we find that only one out of
the three theories is consistent with the principles of effective
field theory~(EFT).  

There is a notable example in which higher-derivative operators such
as $\mathcal{L}_{(1)}$ have
been computed from radiative corrections in a manifest off-shell scheme, namely the effective one-loop superspace Lagrangian of 
the Wess-Zumino model \cite{Buchbinder:1994iw, Pickering:1996he,Kuzenko:2014ypa}. These references focused
purely on those higher-derivative operators that contribute to the
scalar potential and in \cite{Kuzenko:2014ypa} an infinite 
tower of such higher-derivative operators, denoted as the effective
auxiliary field potential (EAFP), was explicitly computed.  
To lowest order in superspace-derivatives this EAFP coincides with
$\mathcal{L}_{(1)}$ given in
eq.~\eqref{eq:higher_derivative_global_susy}. 
The full non-local EAFP turns out to imply a unique on-shell theory. When truncating this EAFP to 
a finite number of terms,
the truncation naively produces multiple on-shell theories. 
Applying the truncation at higher order even 
increases the number of solutions. However, we will show
that at any order of the truncated EAFP there is a unique Lagrangian which
reproduces the dynamics of  the non-local theory 
at  that order and which
is consistent with the principles of EFT.
The remaining theories can be regarded as artefacts of the truncation of the infinite tower of higher-derivative operators
similar to the emergence of ghosts in truncated theories \cite{Simon:1990hd}.

Apart from addressing this conceptual issue 
we proceed to compute the on-shell Lagrangians 
for models with arbitrarily many chiral superfields 
both in global and local supersymmetry.
In particular we focus on the induced correction to the scalar
potential and analyze the situation where the two-derivative 
theory has a minimum with a flat direction which can (or cannot)
be lifted by the presence of~$\mathcal{L}_{(1)}$.

In the second part of this paper we will purely focus on the 
effective action obtained from type IIB flux compactifications 
on Calabi-Yau orientifolds. The background fluxes are able 
to stabilize the complex structure moduli and the
dilaton~\cite{Giddings:2001yu,Dasgupta:1999ss}.
In contrast, all K\"ahler 
moduli are described at leading order by a no-scale supergravity and thus are flat
directions
of the potential.
Perturbative corrections 
for the K\"ahler moduli are induced from $\alpha'$- and $g_s$-corrections in the ten-dimensional action. 
An important example is the leading order $(\alpha')^3$-correction to the 
K\"ahler potential which is computed by reducing higher-curvature terms in ten dimensions \cite{Becker:2002nn}.
This correction breaks the no-scale property, but by itself  does not 
lead to a stabilization. When non-perturbative effects are taken into 
account scenarios with supersymmetric~\cite{Kachru:2003aw} or
non-supersymmetric minima can be found
\cite{Balasubramanian:2005zx}.\footnote{For reviews on 
moduli stabilization, flux compactifications and de Sitter vacua, see e.g.~\cite{Douglas:2006es,Denef:2008wq,Baumann:2014nda}.} There is an intrinsic merit to demonstrate the existence of various classes of meta-stable de Sitter (dS) vacua as explicitly as possible in well-controlled examples of string compactifications.
Thus, we find it worthwhile to explore further possibilities of moduli
stabilization 
using only fully perturbative and explicitly computable contributions.

It is thus of interest to pursue the question to what extent 
 additional $(\alpha')^3$-corrections of the 
ten-dimensional theory can lead to a stabilization of moduli 
without taking into account non-perturbative effects.
Indeed, there are several such contributions which 
have not been discussed in detail, owing to the fact that the explicit structure of many of these 
terms is still unknown. It turns out that these terms do not correct the K\"ahler potential of the four-dimensional 
action, but instead require the presence of higher-derivative
operators such as $\mathcal{L}_{(1)}$ 
as off-shell completions. 
At this point the results of the first part of the paper can be used
since $\mathcal{L}_{(1)}$ precisely links
four-derivative terms to corrections of the potential. 
By computing the four-derivative terms from the explicitly known $R^4$-terms in ten dimensions \cite{Antoniadis:1997eg,Antoniadis:2003sw},
the correction to the potential $V_{(1)}$ can be indirectly inferred. 
We find 
\begin{equation}\label{V1intro}
 V_{(1)} \sim \frac{\Pi_i \, t^i}{\mathcal{V}^4} \ ,
\end{equation}
where the $t^i$ denote the two-cycle volumes, $\mathcal{V}$ the overall
volume and the $\Pi_i$ are 
topological numbers defined as
\begin{equation}
 \Pi_i = \int c_2 \wedge \hat{D}_i \ .
\end{equation}
They encode information of the second Chern class $c_2$ and $\hat{D}_i$ form 
a basis of $H^{1,1}(M,\mathbb{Z})$.\footnote{We perform this
  computation without determining numerical factors   
and for the simple case of $h^{1,1} = 1$ but argue that $V_{(1)}$
given in \eqref{V1intro} also holds for arbitrary  $h^{1,1}$ as long as $W=W_0 = const$.}

We then proceed to study the minima of $V_{(1)}$ taken 
together with the potential obtained from the $\alpha'$-corrected K\"ahler potential.
We show the existence of a model-independent non-supersymmetric minimum of this potential
where all four-cycle volumes are fixed to values $\tau_i \sim \Pi_i$ for any Calabi-Yau threefold with 
$\chi(M)>0$.\footnote{We estimate the typical size of the $\Pi_i$ for
  a specific Calabi-Yau threefold to be ${\cal O}(10-100)$.} 
This result suggests the existence of many new non-supersymmetric vacua 
within the landscape, where stabilization occurs purely from the leading order $\alpha'$-corrections, 
but a more detailed discussion of all possible $\alpha'$-corrections will be necessary to support 
this. 
Furthermore, the minimum only exists if the overall 
 sign of  $\mathcal{L}_{(1)}$ is negative. 
This sign is universal and does not depend on the choice of the Calabi-Yau.
Unfortunately, determining this sign requires the knowledge of the particular linear combination of all additional 4D higher-derivative operators contributing to the 4D four-derivative kinetic terms.
This is beyond the scope of this paper and we leave it for future work.

This paper is organized as follows. In section \ref{susycase}
we study $\mathcal{L}_{(1)}$ in effective theories with global 
supersymmetry. The conceptual discussion of the on-shell theories 
is performed for theories with a single chiral superfield in 
section \ref{onechiral} and in appendix~\ref{sec:appendix_explicit_solutions},
where we also display the exact solutions for the chiral auxiliary field 
and prove the absence of ghosts. In section \ref{sec:appendix_wess_zumino} 
we illustrate the interpretation of the higher-derivative operators 
and the respective on-shell theories with the one-loop Wess-Zumino model.
In section \ref{multifieldglobal} we then display 
the physical on-shell Lagrangian for arbitrarily many chiral 
superfields and make some statements regarding the 
structure of the resulting minima, providing an explicit
example for the lifting of flat directions in section \ref{oraif}.
In section \ref{sugracase} we show the respective 
Lagrangians for the case of supergravity and again 
discuss the structure of the minima with an explicit 
example in section \ref{sec:susy_flat_dir}.
Finally in section \ref{sec:LVS} we turn to the discussion 
of flux compactifications of Type IIB on Calabi-Yau 
orientifold, where the details of the reduction 
of the curvature-terms in ten dimensions can be found 
in appendix~\ref{app:tensoralphaprime} and 
appendix~\ref{app:geometrytensor}. At the end 
we provide some conclusions in section \ref{sec:conclusions}.
\section{Higher-Derivative Terms in $\mathcal{N}=1$ Supersymmetry}
\label{susycase}
\subsection{Preliminaries}
In this section we consider globally supersymmetric theories with 
$n_c$ chiral superfields $\Phi^i$, \, $i=1,\ldots,n_c$
whose couplings are encoded in a K\"{a}hler potential
$K$, a superpotential $W$ and 
the higher-derivative operator $\mathcal{L}_{(1)}$. 
In the following we adopt the conventions and notation of \cite{Wess:1992cp}.
Thus, the total superspace Lagrangian is of the form\footnote{Here and henceforth we drop brackets, which would indicate explicitly on which fields certain superspace-derivatives act. More precisely this means
 $D^{\alpha} \Phi D_{\alpha} \Phi = (D^{\alpha} \Phi) (D_{\alpha} \Phi)$.}
\begin{equation}\begin{aligned}
\label{eq:higher_derivative_global_sec2}
 \mathcal{L} &=  \mathcal{L}_{(0)} + \mathcal{L}_{(1)} \ ,\\
\textrm{where}\qquad\mathcal{L}_{(0)}&= 
\int d^4 \theta \, K(\Phi,\Phi^{\dagger}) + \int d^2
 \theta \,W (\Phi) + \text{h.c.} \ ,\\
\mathcal{L}_{(1)} &=  
 \tfrac{1}{16} \int d^4 \theta\, T_{ij\bar{k}\bar{l}}(\Phi,\Phi^{\dagger})\, D^{\alpha} \Phi^i D_{\alpha} \Phi^j \bar{D}_{\dot{\alpha}} \Phi^{\dagger \bar{k}} \bar{D}^{\dot{\alpha}} \Phi^{\dagger \bar{l}} \ .
\end{aligned}\end{equation}
In the spirit of \cite{Koehn:2012ar} we allow for an arbitrary
hermitian four-tensor superfield $T_{ij\bar{k}\bar{l}}(\Phi,\Phi^{\dagger})$  
which we assume to depend only 
$\Phi$ and $\Phi^{\dagger}$ but
not on any derivative.\footnote{Note that if $T$ 
  would depend on space-time or superspace-derivatives of the
  chiral multiplets the resulting theory would either involve 
more than four derivatives for the component fields and/or 
not correct the scalar potential.} 
We will often refer to this mass dimension $-4$ quantity, respectively its scalar component as 
coupling tensor. From the structure of $\mathcal{L}_{(1)}$ one infers the symmetry properties
\begin{equation}
 T_{ij\bar{k}\bar{l}} = T_{ji\bar{k}\bar{l}} = T_{ji\bar{l}\bar{k}}\ .
\end{equation}
In order to obtain the component expression of $\mathcal{L}$ we use
the well known $\theta$-expansion of the chiral superfields 
\begin{equation}
 \Phi^i = A^i + \sqrt{2} \theta \psi^i + \theta^2 F^i + i \theta
 \sigma^{\mu} \bar{\theta} \partial_\mu A^i - \tfrac{i}{\sqrt{2}}
 \theta \theta \partial_\mu \psi^i \sigma^{\mu} \bar{\theta} + \frac{1}{4} \theta^2 \bar{\theta}^2 \Box A^i \;\;,
\end{equation}
where $A^i$ are scalars, $\psi^i$ chiral fermions and $F^i$  
auxiliary components. From the form of the superspace derivatives 
\begin{equation}
 D_{\alpha} = \frac{\partial}{\partial \theta^{\alpha}} + i \sigma_{\alpha \dot{\alpha}}^{\mu} \bar{\theta}^{\dot{\alpha}} \frac{\partial}{\partial x^{\mu}} \qquad\text{and} \qquad\bar{D}_{\dot{\alpha}} = -\frac{\partial}{\partial \bar{\theta}^{\dot{\alpha}}} -i \theta^{\alpha}  \sigma_{\alpha \dot{\alpha}}^{\mu}  \frac{\partial}{\partial x^{\mu}}\ ,
\end{equation}
one finds that the bosonic part of $\mathcal{L}_{(1)}$
only has a contribution at order 
$\theta^2 \bar{\theta}^2$ which is given by
\begin{equation}\begin{aligned}\label{bosonicpart}
 &T_{ij\bar{k}\bar{l}}(\Phi,\Phi^{\dagger}) D^{\alpha} \Phi^i D_{\alpha} \Phi^{j} \bar{D}_{\dot{\alpha}} \Phi^{\dagger \bar{k}} \bar{D}^{\dot{\alpha}} \Phi^{\dagger \bar{l}}|_{\text{bos}} =\\
 & 16  T_{ij\bar{k}\bar{l}}(A,\bar A)\left[ (\partial_\mu A^i  \partial^\mu A^{j})( \partial_\nu \bar{A}^{\bar{k}} \partial^\nu \bar{A}^{\bar{l}})  - 2 F^i \bar{F}^{\bar{k}} (\partial_\mu A^{j} \partial^\mu \bar{A}^{\bar{l}})+ F^i F^j\bar{F}^{\bar{k}} \bar{F}^{\bar{l}} \right] \theta^2 \bar{\theta}^2\ .
\end{aligned}\end{equation}
Performing the $\theta$ integration in eq.~\eqref{eq:higher_derivative_global_sec2} one obtains the Lagrangian
\begin{equation}\begin{aligned}
\label{Lmulti}
 \mathcal{L}_{\text{bos}} &= - G_{i\bar{j}} \, \partial_{\mu} A^i \partial^{\mu} \bar{A}^{\bar{j}} +  G_{i\bar{j}} \, F^i \bar{F}^{\bar{j}}  + F^i \, W_{,i} + \bar{F}^{\bar{i}} \, \bar{W}_{,\bar{i}} \\
 & \;\;\;\;\; +  T_{ij \bar{k} \bar{l}} (A, \bar{A}) \left[ (\partial_{\mu} A^i  \partial^{\mu} A^{j})( \partial_{\nu} \bar{A}^{\bar{k}}  \partial^{\nu} \bar{A}^{\bar{l}})  - 2 F^i \bar{F}^{\bar{k}} (\partial_{\mu} A^{j} \partial^{\mu} \bar{A}^{\bar{l}})+ F^i F^j\bar{F}^{\bar{k}} \bar{F}^{\bar{l}} \right] \ ,
\end{aligned}\end{equation}
where $G_{i\bar{j}}=\partial_i\partial_{\bar{j}}K$ and $W_{,i}$ denotes the holomorphic
derivative of the superpotential.
We indeed see that no derivative terms for $F^i$ appear and, thus, 
their equations of motion stay algebraic such that the $F^i$ remain non-propagating auxiliary fields.
However, $\mathcal{L}_{\text{bos}}$ contains quartic terms in the $F^i$ 
which lead to cubic contributions to the bosonic part of the respective equations of motion 
\begin{equation}
\label{eq:eom_auxiliary_field_multi}
 G_{i\bar{k}} F^i + \bar{W}_{,\bar{k}} + 2 F^i (F^j \bar{F}^{\bar{l}}
- \partial_\mu A^j \partial^\mu \bar{A}^{\bar{l}}) T_{ij \bar{k} \bar{l}} = 0 \ .
\end{equation}
Determining all solutions to this equation in all generality is a delicate 
task and therefore we first turn to a theory with a single chiral multiplet where we can
solve the cubic equation \eqref{eq:eom_auxiliary_field_multi} exactly.
\subsection{Theory with one Chiral Multiplet}\label{onechiral}
For one chiral multiplet eqs.~\eqref{eq:eom_auxiliary_field_multi} reduce to 
\begin{equation}
\label{eq:eom_auxiliary_field}
 G_{A \bar{A}} \,\bar{F} + W_{,A} + 2
 T \bar{F} \left(\lvert F \lvert^2  - \partial_{\mu} A \partial^{\mu} \bar{A} \right) = 0\ ,
\end{equation}
where we defined $T = T_{AA \bar{A} \bar{A}}$ for brevity. 
In appendix~\ref{sec:full_one_d_sol} we solve 
eq.~\eqref{eq:eom_auxiliary_field} exactly and show that 
depending on $T$ and the specific region in the 
phase space of $A$ one or three solutions for $F$ exist.
Expanding the solutions for small $T$ and inserting into
eq.~\eqref{Lmulti} keeping only the leading terms
one obtains, in the case where all three solutions exist, 
the following three Lagrangians
\begin{equation}\begin{aligned}
 \label{eq:analytic_branch}
 \mathcal{L}_{F_1} & = -G_{A \bar{A}}\left(1+ 2 \hat T V_{(0)} \right)\partial_{\mu} A \partial^{\mu} \bar{A} 
+ \hat{T} G_{A \bar{A}}^2 (\partial_{\mu} A  \partial^{\mu} A )( \partial_{\nu} \bar{A} \partial^{\nu} \bar{A}) \\
 & \quad - V_{(0)} + \hat T  V_{(0)}^2  + \mathcal{O}(\hat T^2) \ , \\
 \mathcal{L}_{F_{2,3}} &= - \tfrac{1}{4}{\hat T}^{-1} + \tfrac{1}{2} V_{(0)} + \mathcal{O}( \hat T^{1/2} ) \ ,
\end{aligned}\end{equation}
where for convenience we defined $\hat T = T G_{A\bar A}^{-2}$ and $V_{(0)}=G^{A \bar{A}} \lvert W_{,A} \lvert^2$ is the scalar potential
of $\mathcal{L}_0$.\footnote{Expanding eq.~\eqref{eq:sol_reg3} one observes that 
$\hat{T}$ is the correct expansion parameter only for $\mathcal{L}_{F_1}$, 
while for $\mathcal{L}_{F_{2,3}}$ it is $\sqrt{\hat{T}}$. Note that at the displayed order in $\sqrt{\hat T}$ the solutions $F_2$ and $F_3$ 
induce the same Lagrangian while at higher order we find $\mathcal{L}_{F_2} \neq \mathcal{L}_{F_{3}}$.} In the following we will sometimes refer to the 
individual Lagrangians in eq.~\eqref{eq:analytic_branch} as branches. 
We observe that $\mathcal{L}_{F_1}$ is analytic in $\hat{T}$ and reproduces $\mathcal{L}_0$ at leading order. At 
linear order in $\hat{T}$ it induces a correction to the kinetic energy, which is proportional to $V_{(0)}$, as well as 
to the potential, proportional to $V_{(0)}^{2}$. $\mathcal{L}_{F_{2,3}}$ on the other hand 
have a pole-like term in $\hat T$ and at order $\hat{T}^0$ only have a contribution to the potential, which differs 
from $V_{(0)}$ by a factor $-1/2$.

In summary the theory defined by 
\eqref{eq:higher_derivative_global_sec2}
can lead to three different and independent on-shell Lagrangians. However, a multiplet of theories is dissatisfying, since it
predicts several inequivalent evolutions of fields for a given set of
initial data. Furthermore, suppose we include additional off-shell higher-derivative operators with more than four superspace-derivatives
then the equations of motion for the chiral auxiliaries admit more than three solutions, rendering the problem even more severe. 
Let us now argue how to resolve this issue in the context of an effective field theory. 

When performing the limit $T\to0$ in the off-shell Lagrangian 
given in eq.~\eqref{eq:higher_derivative_global_sec2} we recover 
the ordinary, two-derivative theory $\mathcal{L}_{(0)}$. For consistency this should 
also hold in the on-shell theories given in eq.~\eqref{eq:analytic_branch}.
For example suppose that the higher-derivative operator 
arises by integrating out massive states associated with a mass scale $M$ from a UV theory. 
Then to lowest order in fields one has $T \sim M^{-4}$ and hence the operator should decouple 
as $M$ becomes large compared to the masses of the light states
as dictated by the decoupling principle, see for instance \cite{Burgess:2007pt}.
We see that $\mathcal{L}_{F_1}$ given in \eqref{eq:analytic_branch} is
analytic in $T$, while $\mathcal{L}_{F_{2,3}}$
contain a non-analytic part and thus violate the decoupling limit. 
Based on this observation we propose to regard only $\mathcal{L}_{F_1}$ 
as the physical on-shell Lagrangian since it is the unique Lagrangian compatible 
with the principles of effective field theory. 
We will substantiate this proposition with the example of 
the effective one-loop Wess-Zumino model in the next section.
Notably we will show that the non-analytic theories not only 
fail to obey the decoupling limit, but furthermore 
are incapable of reproducing the on-shell Lagrangian 
of the full, non-local theory. To some extent this is already 
visible in eq.~\eqref{eq:analytic_branch}. More precisely 
the non-analytic branches fail to reproduce the terms in $\mathcal{L}_{(0)}$. 
In fact they neither include the kinetic terms nor the scalar potential of $\mathcal{L}_{(0)}$.
On the other hand the $\mathcal{O}(T^0)$ contributions in $\mathcal{L}_{F_1}$  
exactly coincide with the terms in $\mathcal{L}_{(0)}$. 
In summary, this observation and the results of the next section 
suggest that the non-analytic solutions should be regarded as mere 
artefacts of the truncation of an infinite sum of higher-derivatives. 
Note that the above observation is reminiscent of the discussion of theories with 
higher-derivative terms in the equations of motion where 
ghost-like degrees of freedom emerge. Similarly the ghosts arise from truncating an infinite
series of higher-derivative terms to a finite sum and violate EFT-reasoning in as much as the inclusion of
higher order operators should merely induce a small correction to the
dynamics of some IR-Lagrangian. A ghost-free theory can then be obtained 
by demanding analyticity of the solutions to the equations of motion 
in EFT-control parameters \cite{Jaen:1986xl, Simon:1990hd}, identical 
to our reasoning above.

In the rest of this paper we will therefore only discuss the 
analytic theory. Furthermore, recall that besides the operator in 
eq.~\eqref{eq:higher_derivative_global_sec2} 
superspace higher-derivative terms with more than four superspace-derivatives
exist and they contribute higher polynomial powers of the auxiliary field to the 
Lagrangian (next section we display the one-loop Wess-Zumino model as an explicit example where infinitely many superspace-derivative operators are present). These operators are further mass-suppressed and hence modify the equations of motion for 
the auxiliary fields at order $\mathcal{O}(T^2)$.\footnote{They might also induce modifications at order $\mathcal{O}(T^{3/2})$.} 
This implies that without including such higher-derivative terms into 
the superspace Lagrangian, we can trust the resulting on-shell Lagrangian only 
up to linear order in $T$.\footnote{In models, where the EAFP is solely given in terms of the four-derivative operator, it is sensible to regard the full solution for $F^i$ and the respective Lagrangians along the lines of appendix~\ref{sec:appendix_explicit_solutions}.}
Fortunately this greatly simplifies the structure of the on-shell 
Lagrangian and makes a proper discussion of 
the multi-field case feasible.

To conclude this section let us describe why the theory is free of ghosts. 
The absence of ghosts is not immediately clear, but can be understood with the exact solution 
for the auxiliary field at hand. The sign of the ordinary kinetic term is affected by the presence 
of the higher-derivative operator through eq.~\eqref{eq:eom_auxiliary_field}. 
In appendix~\ref{sec:appendix_ghosts} the absence of ghosts is explicitly 
demonstrated for the theory obtained by solving eq.~\eqref{eq:eom_auxiliary_field} 
exactly and reinserting the result into~eq.~\eqref{Lmulti}. Nevertheless, one might still 
worry about the sign of the ordinary kinetic term in the truncated theory after 
inspection of eq.~\eqref{eq:analytic_branch}. More precisely one finds that the theory becomes 
ghost-like once $ \hat T V_{(0)} \sim -1$. However, in that 
regime we cannot trust our truncation at linear order in $T$ any longer 
as we illustrate in appendix~\ref{sec:appendix_explicit_solutions}. 
In other words, studying the exact solutions of eq.~\eqref{eq:eom_auxiliary_field} 
shows that if $ \hat T V_{(0)} \sim -1$, 
the analytic solution ceases to exist and one enters a regime, in 
which only non-perturbative solutions can be found. To summarize, 
the analytic theory breaks down before it would become ghostlike.
\subsection{One-loop Wess-Zumino Model}
\label{sec:appendix_wess_zumino}
After the general discussion of the previous section let us now turn to an explicit example, where 
the truncation of the infinite sum of higher-derivatives and the structure of the equations of motion 
for the auxiliary field can be explicitly studied. This example is given by 
the one-loop Wess-Zumino model in superspace, for which the full, non-local effective auxiliary field potential
(EAFP) was recently computed in \cite{Kuzenko:2014ypa} following up on earlier works \cite{Buchbinder:1994iw,Pickering:1996he}. 
More precisely the model consists of a single chiral superfield $\Phi$ with K\"ahler potential and superpotential of the form 
\begin{equation}
 K = \Phi \Phi^\dagger \ , \qquad W=\frac12 m\Phi^2+\frac16\lambda\Phi^3 \ .
\end{equation}
According to \cite{Kuzenko:2014ypa} the only contributions to the effective superspace potential at one-loop
come from corrections to the K\"{a}hler potential as well as an EAFP, which we denote as $\mathbb{F}$.
More precisely it consists of an infinite tower of higher-derivatives of the form
\begin{equation}\label{WZEAFP1}
 \mathbb{F} = \int\mathrm{d}^4 \theta\ \frac{D \Psi D \Psi \bar{D} \Psi^{\dagger} \bar{D} \Psi^{\dagger}}{(\Psi \Psi^{\dagger})^2} \ G\left(\frac{D^2 \Psi \bar{D}^2 \Psi^{\dagger}}{(\Psi \Psi^{\dagger})^2} \right) \;\;\;,
\end{equation}
where $\Psi = m + \lambda \Phi = W^{''}$ and $G$ is a known real-valued
analytic function with non-vanishing coefficients in the respective 
series expansion at all orders \cite{Kuzenko:2014ypa}. The lowest order contribution arises from 
the constant term in the series expansion of $G$ and comparing with
\eqref{eq:higher_derivative_global_sec2} we have 
\begin{equation}
T \sim |W^{''}|^{-4}\ .
\end{equation}
Expanding $T$ as a geometric series, we 
identify that to lowest order we have $T \sim m^{-4}$.

Let us now proceed by performing the superspace integration in eq.~\eqref{WZEAFP1}. 
From eq.~\eqref{bosonicpart} we infer that the bosonic part of
the superfield multiplying $G$ has only a $\theta^2 \bar{\theta}^2$
contribution and hence the remaining superfields have to be evaluated at
their scalar component.
This yields
\begin{equation}\label{EAFPWZ}
 \mathbb{F}_{\text{bos}} = \frac{(D \Psi D \Psi \bar{D} \Psi^{\dagger} \bar{D} \Psi^{\dagger})\lvert_{\theta^4}}{\lvert m + \lambda A \lvert^4} \, G\left(\frac{\lvert \lambda F \lvert^2}{\lvert m + \lambda A \lvert^4} \right) \ .
\end{equation}
For simplicity let us set $\lambda = 1$ from now on. $\mathbb{F}_{\text{bos}}$ displays 
an infinite sum in the auxiliary field $F$ and $\bar{F}$. Additional powers of the 
auxiliary field are in a one-to-one correspondence with additional powers of 
superspace-derivatives. We can identify 
\begin{equation}
 \epsilon \equiv \lvert m + A \lvert^{-4}
\end{equation}
as the parameter controlling the infinite series of higher-derivatives and
powers of the auxiliary field, respectively. We immediately observe that 
eq.~\eqref{EAFPWZ} comprises an analytic function in $\epsilon$.
Using the full (and explicitly known) function $G$ it can be numerically 
shown that the solution to the equations of motion for $F$ derived 
from the standard Lagrangian plus $\mathbb{F}_{\text{bos}}$ is unique 
and analytic in $\epsilon$.

The non-local theory with $\mathbb{F}$ in eq.~\eqref{EAFPWZ} can be regarded as 
a UV-theory for a local theory after truncating the infinite 
sum of higher-derivatives to a finite sum. For the purpose of obtaining 
a local theory also the control parameter $\epsilon$ has to be 
truncated. However, we omit this here, as it does not provide additional 
insight into the structure of the series in higher-derivatives.

It is interesting to discuss the
equations of motion for the auxiliary field once the theory is
truncated at a given order in $\epsilon$. In the following let $G_n$ denote the truncation of the series expansion of $G$ at order $n$. If we truncate $G$ at $\mathcal{O}(\epsilon)$, the discussion reduces to the familiar cubic in eq.~\eqref{eq:eom_auxiliary_field}, which admits only one analytic solution. For arbitrary $n$ the contribution of eq.~\eqref{EAFPWZ} to the scalar potential reads
\begin{equation}
\mathbb{F}_{\text{bos}} \sim \epsilon \lvert F \lvert^4 G_n(\epsilon
\lvert F \lvert^2)\ .
\end{equation}
Taking into account the remaining, ordinary terms in the Lagrangian, i.e.~$\mathcal{L}_{(0)}$ in eq.~\eqref{eq:higher_derivative_global_sec2}, the equation of motion for $F$ reads
\begin{equation}
\label{eq:cubic_truncated_WZ}
 F + \bar{W'} + 2 \epsilon F \lvert F \lvert^2 G_n(\epsilon
 \lvert F \lvert^2) + \epsilon^2 F \lvert F \lvert^4 G_n'(\epsilon
 \lvert F \lvert^2) = 0\ ,
\end{equation}
where we only took into account terms that contribute to the scalar potential. $G_n$ induces monomials in $\lvert F \lvert^2$ up to degree $n$ and, hence, eq.~\eqref{eq:cubic_truncated_WZ} admits up to $(2n+3)$ independent solutions. In other words the number of solutions is increasing with the order of the truncation. To solve eq.~\eqref{eq:cubic_truncated_WZ} we first redefine the auxiliary field via
\begin{equation}
 F = \bar{W'} f \ .
\end{equation}
Inserted into eq.~\eqref{eq:cubic_truncated_WZ} one observes that $f$ has to be real and, hence, eq.~\eqref{eq:cubic_truncated_WZ} reduces to
\begin{equation}\label{eomtrunred}
 f + 1 + 2 \epsilon f^3 \lvert W' \lvert^2 G_n(\epsilon f^2 \lvert W' \lvert^2) + \epsilon^2 f^5 \lvert W' \lvert^4 G_n'(\epsilon f^2 \lvert W' \lvert^2) = 0 \ .
\end{equation}
We make an ansatz of the form
\begin{equation}\label{ansatzexp}
 f = \sum_{i=-1}^{\infty} \epsilon^{i/2} f_i \ ,
\end{equation}
such that eq.~\eqref{eomtrunred} at lowest order in $\epsilon$ reads
\begin{equation}
\label{equation_fminusone}
 f_{-1} + f_{-1}^3 \lvert W' \lvert^2 G_n(f_{-1}^2 \lvert W' \lvert^2) +  f_{-1}^5 \lvert W' \lvert^4 G_n'( f_{-1}^2 \lvert W' \lvert^2) = 0 \ .
\end{equation}
Since $G_n$ is a polynomial of degree $n$ with non-vanishing
coefficients we see that only the branch given by $f_{-1} = 0$ is
analytic. All other solutions, which are defined at lowest order by the remaining $2n+2$ solutions of eq.~\eqref{equation_fminusone} and necessarily fulfill $f_{-1} \neq 0$, 
are non-analytic in $\epsilon$ for any $n$. 

In effective field theory one generally expects to be able to compute observables with 
higher precision by including more and more operators. Indeed since
the unique solution of the non-local theory was analytic, the analytic solution
of the truncated theory is able to reproduce the Lagrangian of the non-local theory at order $\epsilon^{n+1}$
and, thus, mimics the non-local theory with better precision for larger $n$. 
However, regardless of the order of the truncation the non-analytic theories fail to reproduce 
the non-local theory to that specific order. One can explicitly check this for the first components
in the expansion in eq.~\eqref{ansatzexp}. At lowest order this was also already visible 
in eq.~\eqref{eq:analytic_branch}. 

It is worth noting that the existence of a unique analytic solution 
for $F$ in the truncated theory does not depend on the details of 
the $\mathbb{F}$, but we expect it to hold in general as long as the coefficient of 
the $\lvert F \lvert^2$ term in the Lagrangian is non-vanishing. 
Indeed the EAFP is correcting the Lagrangian by at least cubic powers of $F$ 
and $\bar{F}$ \cite{Buchbinder:1994iw} so that one would always 
expect the analytic solution to be unique.

After the above conceptual discussion we can now proceed to study
theories with more than one chiral multiplet. 
\subsection{Multi-Field Case and Analysis of Scalar Potential}\label{multifieldglobal}
Given the results of the previous sections we constrain the discussion of the multi-field case to the analytic solution of eq.~\eqref{eq:eom_auxiliary_field_multi}. 
Solving eq.~\eqref{eq:eom_auxiliary_field_multi} using perturbation theory yields at linear order in $T$
\begin{equation}\begin{aligned}
\label{eq:f-term_global}
 F^i  =& F_{(0)}^i + F_{(1)}^i\ ,\qquad \textrm{where}\qquad   F_{(0)}^i =  -  G^{i\bar{l}} \,\bar{W}_{,\bar{l}} \ , \\
 F_{(1)}^i =& 2 \,T^{\bar{k}\bar{l} ij}\,\bar{W}_{,\bar{k}} \, \bar{W}_{,\bar{l}} \, W_{,j} 
 -2 T^{\bar{k}} {}_j {}^i {}_{\bar{l}} \,(\partial_\mu A^j  \partial^\mu \bar{A}^{\bar{l}})\,\bar{W}_{,\bar{k}} \ .
\end{aligned}\end{equation}
Insertion of the auxiliary field into the Lagrangian in eq.~\eqref{Lmulti} yields
\begin{equation}\label{globalL}
\mathcal{L}_{\text{bos}} = - \left( G_{i \bar{k}} + 2 T^{\bar{l}} {}_i {}^j {}_{\bar{k}} \,W_{,j} \,  \bar{W}_{,\bar{l}}  \right) \partial_\mu A^i  \partial^\mu \bar{A}^{\bar{k}} +  T_{ij \bar{k} \bar{l}} \,(\partial_\mu A^i  \partial^\mu A^j)(\partial_\mu \bar{A}^{\bar{k}}  \partial^\mu \bar{A}^{\bar{l}}) - V(A,\bar{A}) \ .
\end{equation}
The resulting scalar potential at linear order in $T$ reads
\begin{equation}
\label{potglobal}
 V =  V_{(0)} + V_{(1)}\ ,\qquad \textrm{where}\qquad
V_{(0)} = G^{i \bar{j}} W_{,i}  \bar{W}_{,\bar{j}}\ ,\quad
V_{(1)} = - T^{ij \bar{k} \bar{l}}  W_{,i}  W_{,j}  \bar{W}_{,\bar{k}}  \bar{W}_{,\bar{l}} \ .
\end{equation}
Before we analyse this potential, let us make a comment regarding the ordinary kinetic term in the Lagrangian in eq.~\eqref{globalL}. The metric multiplying the kinetic term is corrected by
\begin{equation}\label{deltaGik}
 \delta G_{i \bar{k}} = 2 T^{\bar{l}} {}_i {}^j {}_{\bar{k}} \,W_{,j} \,  \bar{W}_{,\bar{l}} \ .
\end{equation}
In general it is not possible to absorb the correction in eq.~\eqref{deltaGik} by performing a change of coordinates in field-space and, hence, the metric multiplying the kinetic term in eq.~\eqref{globalL} is in general not a K\"ahler metric.\footnote{However, this metric is still hermitian.} For the following special form of the coupling tensor 
\begin{equation}
\label{eq:coupling_tensor_metric}
 T_{ij \bar{k}\bar{l}} = \frac{T}{2} \left( G_{i\bar{k}} G_{j\bar{l}} + G_{i\bar{l}}G_{j\bar{k}} \right)\; ,
\end{equation}
with $T=const.$ this was demonstrated explicitly in \cite{Cecotti:1986jy}.

Since the supersymmetry transformations of the chiral multiplets do not change,
the order parameter for supersymmetry breaking continues to be
$\langle F^i\rangle$. Therefore the supersymmetric minima of
$V$ are found at 
\begin{equation}\label{susymin}
\langle F^i\rangle = 0 \ .
\end{equation}
From eq.~\eqref{eq:eom_auxiliary_field_multi} we see that the supersymmetric locus in field space $\langle A^i\rangle$ which solves
\eqref{susymin} is determined by $\langle F^i_{(0)}\rangle = \langle W_{,i}\rangle= 0$ and, thus, is not
corrected by the presence of the higher-derivative terms under the condition that $T$ is non-singular.\footnote{This can also be inferred from eq.~\eqref{eq:f-term_global}. However, care must be taken as eq.~\eqref{eq:f-term_global} suggests that up to two additional solutions to eq.~\eqref{susymin} exist for which $\langle W_{,i}\rangle \neq 0$. Yet these would be due to a non-trivial cancellation between $F^i_{(0)}$ and $F^i_{(1)}$ that will be spoiled once higher order corrections in $T$ to $F^i$ are considered. More precisely these solutions would only exist because we truncate the auxiliary field at a certain order 
and are, thus, artefacts of this truncation.}
Indeed it was shown that for arbitrary higher-derivative theories the structure of the supersymmetric 
vacua is unchanged \cite{Cecotti:1986jy}. In particular this implies that any flat direction of $V_{(0)}$ is not lifted.

If supersymmetry 
is broken by some 
$\langle F^i_{(0)}\rangle \neq 0$ the higher-derivative correction 
can become important. Still 
 $V_{(1)}$ is a perturbation of $V_{(0)} $
and therefore the minimum 
$\langle A^i_{(0)}\rangle$ of $V_{(0)} $
will at best be shifted to a nearby field value 
$\langle A^i_{(0)}\rangle\to 
\langle A^i_{(0)}\rangle + \langle \delta A^i\rangle$.
However, if the non-supersymmetric minimum of $V_{(0)} $
has a flat direction the contribution from $V_{(1)}$ becomes the leading
term in this direction and may lift its flatness. A possible 
exception to this occurs when the flatness is due to a symmetry, such as 
a perturbatively unbroken shift-symmetry. Further exceptions are 
models in which supersymmetry breaking occurs due to a 
spontaneously broken R-symmetry \cite{Nelson:1993nf}. 
In this case there always exists a flat direction, the R-axion, associated with the 
Goldstone boson of the broken R-symmetry. Here the existence 
of higher-derivative corrections does not lift the flatness.

If the flatness is lifted, then depending on the structure and sign of 
$T$ the flat direction can be stabilized or destabilized. 
It is difficult to make a general statement, and in the end
a case-by-case analysis is necessary.
Nevertheless, before we proceed, let us offer some general observations.

A (real) flat direction $\phi$ is characterized by the fact the all
$\phi$-derivatives of $V$ vanish in the background, or in other words
\begin{equation}\label{Vflat}
\langle \partial_\phi^n  V\rangle =0\ , \quad \forall
n\in \mathbb{N}\ .
\end{equation}
Let us assume that $V_{(0)}$ has a flat direction and thus satisfies
\eqref{Vflat}. A special (and simple) case of this situation
is that $V_{(0)}$ does not
depend on $\phi$ at all, i.e.\  
$\partial_\phi^n V_{(0)} \equiv0, \forall n$.
In this case the flat direction is lifted for generic $T$
but preserved if $T$ is also independent of $\phi$. 
A slight generalization occurs when $W_{,i}$ and only the matrix element
of $G^{i \bar{j}}$ in the direction of the supersymmetry breaking $F$-term,
say $F^0$, are independent of $\phi$. In this case the flat direction
is preserved if also $T^{00\bar 0\bar 0}$ is independent of $\phi$. 
As a final example let us discuss a specific form of the coupling tensor given in eq.~\eqref{eq:coupling_tensor_metric}.
In this case we have $V_{(1)} = - T V_{(0)}^2$ and thus any flat direction of 
$V_{(0)}$ remains flat with respect to $V_{(1)}$, given that the scalar
function $T$ does not depend upon it.
\subsection{Example: O'Raifeartaigh Model}\label{oraif}
For concreteness let us discuss a specific example of a model with flat directions within non-supersymmetric vacua. The simplest case is given by the O'Raifeartaigh model. This is defined via a K\"{a}hler and superpotential, which read
\begin{equation}
\label{oraifeartaigh}
 K = \lvert A_0 \lvert^2 + \lvert A_1 \lvert^2 + \lvert A_2 \lvert^2 \ , \qquad W = \lambda A_0 + m A_1 A_2 + Y A_0 A_1^2 \ .
\end{equation}
Here $\lambda, m , Y$ are real parameters such that $m^2 > 2 \lambda Y$. The resulting potential is minimized at $\langle A_1 \rangle = \langle A_2 \rangle = 0$ leaving $A_0$ unfixed. Since $\langle V_0 \rangle = \langle \lvert F_0 \lvert^2 \rangle = \lambda^2$, supersymmetry is broken in the vacuum. Eq.~\eqref{oraifeartaigh} has a $\mathbb{Z}_2$-symmetry in $A_1$ and $A_2$ and furthermore an R-symmetry, if we assign R-charges as follows
\begin{equation}
 R(A_0) = R(A_2) = 2 \ ,\qquad R(A_1) = 0 \ .
\end{equation}
For the continuum of vacua labeled by $\langle A_0 \rangle$ there exists one vacuum, namely $\langle A_0 \rangle = 0$, in which the R-symmetry is not spontaneously broken. Thus, the O'Raifeartaigh model is an exception to the generic expectation that supersymmetry breaking occurs due to R-symmetry breaking in models, which reduce to Wess-Zumino models in the low energy regime and respect the principles of EFT \cite{Nelson:1993nf}.

Let us proceed by switching on the higher-derivative operator. We consider vacua in which $\langle A_1 \rangle = \langle A_2 \rangle = 0$ as in the ordinary theory. The respective potential at the point $A_1 = A_2 = 0$ is extremized, if the following holds
\begin{equation}
\label{minimum_oraif}
 \partial_{i} V = - T^{00\bar{0} \bar{0}}_{,i} \lambda^4 - 2m\lambda^3(1-\delta_{i,0})(T^{i0\bar{0} \bar{0}} + T^{00\bar{i} \bar{0}}) = 0 \ .
\end{equation}
We see that the flatness of $A_0$ is lifted, if certain components of the tensor require a specific value for extremization. 

Inspecting eq.~\eqref{eq:higher_derivative_global_sec2} we find that the higher-derivative Lagrangian is R-symmetric, if 
\begin{equation}
 R(T_{ij\bar{k}\bar{l}})=0 \ .
\end{equation}
The most general coupling tensor at quadratic order in fields respecting the $\mathbb{Z}_2$- and R-symmetry is given by
\begin{equation}\begin{aligned}
T = \,T_{(0)} + T_{(1)} \lvert A_0 \lvert^2 + T_{(2)} \lvert A_1 \lvert^2 + T_{(3)} \lvert A_2 \lvert^2   + T_{(4)}(A_1^2 + \bar{A}_1^2) \ .
\end{aligned}\end{equation}
For simplicity we suppressed the tensor indices of $T$ and $T_{(0)},\dots,T_{(4)}$ here. From eq.~\eqref{minimum_oraif} we see that $A_0$ is fixed in the minimum to the value $\langle A_0 \rangle = 0$, in which the R-symmetry is preserved, unless the following couplings vanish
\begin{equation}
 T_{(1)}^{00\bar{0} \bar{0}} = T_{(1)}^{10\bar{0} \bar{0}} + T_{(2)}^{00\bar{1} \bar{0}} = T_{(1)}^{20\bar{0} \bar{0}} + T_{(1)}^{00\bar{2} \bar{0}} = 0 \ .
\end{equation}
In a generic effective field theory there is no reason why these couplings could be zero and so one concludes that indeed $A_0$ is fixed. Note furthermore that if the R-symmetry would have been broken in the minimum, then a flat direction associated with the respective Goldstone boson would have persisted. Finally, note that the flatness of $A_0$ can also be lifted by including higher-dimensional operators into the K\"ahler- or superpotential.
%
\section{Higher-Derivative Terms in $\mathcal{N}=1$ Supergravity}
\label{sugracase}
\subsection{Preliminaries}
Let us now couple the theory specified in
\eqref{eq:higher_derivative_global_sec2}
to supergravity. We will only reproduce the essential steps here
and refer the reader for a detailed derivation to the original
paper \cite{Koehn:2012ar}.
Without any higher-derivative operator the  Lagrangian is given by \cite{Wess:1992cp}
\begin{equation}
\label{eq:lagrangian_ordinary_sugra}
 \mathcal{L}_{(0)} = \int \mathrm{d}^2 \Theta \,2 \mathcal{E}
 \left[\tfrac{3}{8}(\bar{\mathcal{D}}^2-8R)\mathrm{e}^{-K(\Phi^i,\Phi^{\dagger j})/3}+W(\Phi_i)
 \right] + h.c.\ ,
\end{equation}
where $\mathcal{E}$ denotes the chiral density, $R$ the curvature
superfield and $\bar{\mathcal{D}}^2 =
\bar{\mathcal{D}}_{\dot{\alpha}}
\bar{\mathcal{D}}^{\dot{\alpha}}$ with
$\bar{\mathcal{D}}_{\dot{\alpha}}$
being  the covariant spinorial derivative. To obtain the
Einstein-frame Lagrangian for the scalar fields $A_i$, it is necessary
to perform a Weyl transformation of the vielbein and successively
integrate out all the auxiliary fields. This results in the familiar scalar potential
\begin{equation}\label{Vsugra}
 V_{(0)} = \mathrm{e}^K \left(G^{i \bar{j}} D_i W
    \bar{D}_{\bar{j}} \bar W - 3 \lvert W \lvert^2 \right)\ ,
\end{equation}
where  $D_i W=W_{,i}+K_{,i}W$ is the K\"{a}hler covariant derivative of the
superpotential. 

To couple the higher-derivative operator of
eq.~\eqref{eq:higher_derivative_global_sec2}
to supergravity one can either add the term \cite{Koehn:2012ar}
\begin{equation}
\label{eq:lagrangian_superspace_hd}
 \mathcal{L}_{(1)} = - \frac{1}{64}\int \mathrm{d}^2 \Theta \, \mathcal{E} (\bar{\mathcal{D}}^2-8R) \mathcal{D} \Phi^i \mathcal{D} \Phi^j \bar{\mathcal{D}} \Phi^{\dagger \bar{k}} \bar{\mathcal{D}} \Phi^{\dagger\bar{l}} T_{ij \bar{k}\bar{l}} + h.c.
\end{equation}
to \eqref{eq:lagrangian_ordinary_sugra} 
or  modify the K\"{a}hler potential as\footnote{This type of procedure of coupling a higher-derivative operator to supergravity was also used in \cite{Baumann:2011nm}.}
\begin{equation}
\label{eq:modified_kahler_potential}
 K(\Phi^i,\Phi^{\dagger \bar{j}}) \rightarrow K(\Phi^i,\Phi^{\dagger \bar{j}}) +
 \tfrac{1}{16} T_{ij \bar{k}\bar{l}}\, \mathcal{D} \Phi^i
 \mathcal{D} \Phi^j \bar{\mathcal{D}} \Phi^{\dagger \bar{k}}
 \bar{\mathcal{D}} \Phi^{\dagger\bar{l}} \ .
\end{equation}
 Due to \eqref{bosonicpart} the bosonic Lagrangians obtained by the two methods coincide up to a
K\"{a}hler factor, which can be absorbed in a redefinition of $T$.
Here we assume that $T_{ij\bar{k}\bar{l}}$ only depends on the chiral and anti-chiral 
superfields $\Phi$ and $\Phi^\dagger$ but not on the gravitational multiplet. 

In the Lagrangian $\mathcal{L} =  \mathcal{L}_{(0)} + \mathcal{L}_{(1)}$
one performs the same Weyl-transformation as before and integrates out the auxiliary fields in the
gravitational multiplet. This procedure is not affected by the presence of $\mathcal{L}_{(1)}$. 
One is then left with the Lagrangian 
\cite{Koehn:2012ar}
\begin{equation}\begin{aligned}\label{sugrawf}
\frac{\mathcal{L}_{\text{bos}}}{\sqrt{-g}}  =& -\tfrac{1}{2} \mathcal{R} 
- G_{i \bar{k}} \partial_\mu A^i  \partial^\mu \bar{A}^{\bar{k}} 
+ G_{i \bar{k}} \mathrm{e}^{K/3} F^i \bar{F}^{\bar{k}} 
+ \mathrm{e}^{2K/3} \left[F^i D_i W 
               + \bar{F}^{\bar{k}} \bar{D}_{\bar{k}} \bar W \right] +
 3\mathrm{e}^K \lvert W \lvert^2 \\
  &+  T_{ij \bar{k} \bar{l}} (\partial_\mu A^i  \partial^\mu A^j)(\partial_\nu \bar{A}^{\bar{k}}  \partial^\nu \bar{A}^{\bar{l}}) 
-2 T_{ij \bar{k}\bar{l}} \mathrm{e}^{K/3} F^i \bar{F}^{\bar{k}} (\partial_\mu A^{j} \partial^\mu \bar{A}^{\bar{l}}) 
+ T_{ij \bar{k}\bar{l}} \mathrm{e}^{2K/3} F^i F^j
  \bar{F}^{\bar{k}} \bar{F}^{\bar{l}}\ .
\end{aligned} \end{equation}
The equations of motion for $F^i$ now read
\begin{equation}
\label{eomauxiliary}
 G_{i\bar{k}} F^i + \mathrm{e}^{K/3}\bar{D}_{\bar{k}} \bar{W} + 2 F^i (\mathrm{e}^{K/3}F^j \bar{F}^{\bar{l}}- \partial_\mu A^j  \partial^\mu \bar{A}^{\bar{l}}) T_{ij \bar{k} \bar{l}} = 0 \;\;.
\end{equation}

After the discussion in the previous section we only focus on
the analytic solution of \eqref{eomauxiliary}.\footnote{For the special case in eq.~\eqref{eq:coupling_tensor_metric} we determine the exact analytic solution in appendix~\ref{app:fullsol}.} 
Here it is sufficient to know the auxiliary fields up to linear order in the 
coupling tensor. They read
\begin{equation}\begin{aligned}
\label{eq:f-term}
 F^i  &= F_{(0)}^i + F_{(1)}^i\ ,\qquad
 F_{(0)}^i =  - \mathrm{e}^{K/3} G^{i\bar{k}} \,\bar{D}_{\bar{k}}  \bar{W}, \\
 F_{(1)}^i & = 2 \mathrm{e}^{4K/3}\,T^{\bar{k}\bar{l} ij} \,\bar{D}_{\bar{k}} \bar{W} \, \bar{D}_{\bar{l}} \bar{W} \,D_j W 
 -2 \mathrm{e}^{K/3} T^{\bar{k}} {}_j {}^i {}_{\bar{l}} \,(\partial_\mu A^j  \partial^\mu \bar{A}^{\bar{l}})\,\bar{D}_{\bar{k}}\bar{W} \ .
\end{aligned}\end{equation}
Inserting the above auxiliary field into the Lagrangian in eq.~\eqref{sugrawf} yields
\begin{equation}\begin{aligned}\label{sugraL}
\frac{\mathcal{L}_{\text{bos}}}{\sqrt{-g}}  =& -\tfrac{1}{2} \mathcal{R} 
- \left( G_{i \bar{k}} + 2 \mathrm{e}^K T^{\bar{l}} {}_i {}^j {}_{\bar{k}} \,D_j W \, \bar{D}_{\bar{l}} \bar{W}  \right) \partial_\mu A^i  \partial^\mu \bar{A}^{\bar{k}} \\
 &+  T_{ij \bar{k} \bar{l}} (\partial_\mu A^i  \partial^\mu A^j)(\partial_\nu \bar{A}^{\bar{k}}  \partial^\nu \bar{A}^{\bar{l}}) - V(A,\bar{A}) \ .
\end{aligned} \end{equation}
The scalar potential is corrected as follows
\begin{equation}
\label{eq:corr_pot_sugra}
V =  V_{(0)} + V_{(1)}\ ,
\end{equation}
where $V_{(0)}$ is given in \eqref{Vsugra} while
\begin{equation}\label{V1}
V_{(1)} = -  \mathrm{e}^{2K}T^{\bar{i}\bar{j} kl} \bar{D}_{\bar{i}} \bar{W} \bar{D}_{\bar{j}} \bar{W} D_k W D_l W \ .
\end{equation}
Analogous to eq.~\eqref{globalL} the metric multiplying the ordinary kinetic term receives a correction. From eq.~\eqref{sugraL} we read off its form
\begin{equation}\label{deltaGSUGRA}
 \delta G_{i\bar{k}} = 2\mathrm{e}^K T^{\bar{l}} {}_i {}^j {}_{\bar{k}} \,D_j W \, \bar{D}_{\bar{l}} \bar{W} \ .
\end{equation}
As in the global case this correction in general renders the metric non-K\"ahler.
\subsection{Fate of Flat Directions and Simple No-Scale Examples}
\label{sec:susy_flat_dir}
Let us begin the analysis with the supersymmetric minima
of the potential given in \eqref{Vsugra},\eqref{eq:corr_pot_sugra} 
and \eqref{V1}. $\langle F^i\rangle$ denotes the order parameter for
supersymmetry breaking. Analogous to the discussion with global 
supersymmetry eq.~\eqref{eomauxiliary} implies that unbroken 
supersymmetry imposes the exact same condition as in a standard 
two-derivative supergravity, that is 
\begin{equation}\label{sugramin}
\langle F^i\rangle = \langle D_i W \rangle =0\ ,\qquad
\langle V\rangle = -3\langle \mathrm{e}^K \lvert W \lvert^2 \rangle\ .
\end{equation}
Thus, the location of the supersymmetric minima
in field space are determined by $F^i_{(0)}=0$ and they are unaffected 
by the presence of  $F^i_{(1)}$.
In particular, any flat direction of $V_{(0)}$ is preserved 
by $V_{(1)}$. In addition,
$\langle W\rangle=0$ corresponds to a Minkowski vacuum while
$\langle W\rangle\neq0$ corresponds to an AdS vacuum.

Let us now turn to minima with spontaneously broken supersymmetry.
As in the global case $V_{(1)}$ is considered to be a perturbation of $V_{(0)} $
and the minimum $\langle A^i_{(0)}\rangle$ of $V_{(0)}$ is shifted to a nearby field value 
$\langle A^i_{(0)}\rangle\to \langle A^i_{(0)}\rangle + \langle \delta A^i\rangle$.
Therefore qualitatively nothing changes except for the flat directions. 
Contrary to the case of global supersymmetry in the local case non-trivial models with 
vanishing potential exist. These are the no-scale models. 
The no-scale property is generally expected to be lost when higher-derivative corrections 
are taken into account, thus making it possible to lift flat directions.
In the rest of this section we present a simple example to illustrate 
the fate of flat directions and make a first step towards the 
potential relevance to moduli stabilization.
%
%

More precisely we consider a model specified by a constant superpotential $W(A) = W_0$ and
the K\"{a}hler potential
\begin{equation}
\label{eq:1dtoymodel}
 K(A,\bar{A}) = - p \,\mathrm{ln} (A + \bar{A}) \ , 
\end{equation}
where $p>0$. This $K$ is of the no-scale type in that it satisfies
\begin{equation}
\label{no_scale_type}
 G^{A \bar{A}} K_{,A} K_{,\bar{A}} = p \ .
\end{equation}
In this case $V_{(0)}$ given in \eqref{Vsugra} is positive (negative)
for $p>3\, (p<3)$ and vanishes identically 
for $p=3$. Adding $V_{(1)}$ given in \eqref{V1}
and redefining $\hat{T} = T(A,\bar{A}) G_{A \bar{A}}^{-2}$ one obtains
\begin{equation}
\label{eq:one_dimensional_model_potential}
 V =  V_{(0)} + V_{(1)}= (A+\bar{A})^{-p} (p-3) \lvert W_0 \lvert^2 - \hat{T} (A+\bar{A})^{-2p} p^2 \lvert W_0 \lvert^4 \ .
\end{equation}

For $p=3$ both real and imaginary parts of $A$ are flat directions of $V_{(0)}$.
We see that generically both flat directions are lifted
unless the combination $\hat{T} (A+\bar{A})^{-6}$ is constant in 
$\mathrm{Re}(A)$ and/or $\mathrm{Im}(A)$. For example 
a continuous shift symmetry $A\to A+$ i const.\ which often holds 
perturbatively in string theory would protect the flat direction along
$\mathrm{Im}(A)$ in that $\hat{T}$ could not depend on $\mathrm{Im}(A)$.
In order to say something about the stability, however, one has to make
some assumptions about the functional dependence of $\hat{T}$.

Let us now consider a very simple situation, in which 
the inclusion of $V_{(1)}$ stabilizes a certain 
direction. For instance if $p<3$ and $\hat{T}=$ const.,\footnote{$\hat{T}=$ const.\ can be motivated by the explicit computation of four-derivative terms in \cite{Burton:1989ai}. There the one-loop corrections to the typical no-scale supergravity inspired by the heterotic string were computed.}
  the two terms in
eq.~\eqref{eq:one_dimensional_model_potential} can balance
for $\hat{T}<0$ with a non-supersymmetric AdS minimum at
\begin{equation}
\label{eq:minimum_toy_model}
\langle A + \bar{A}\rangle  = \left(\frac{2p^2}{p-3}\, \hat{T}\, \lvert W_0 \lvert^2 \right)^{1/p} \ , \qquad\textrm{and}\qquad
\langle V \rangle = \frac{(p-3)^2}{4p^2 \hat{T}} < 0\ .
\end{equation}
Furthermore we have to check whether the field-value in eq.~\eqref{eq:minimum_toy_model} is within the regime, where the perturbative solution for the auxiliary field 
converges. An estimate for the boundary between the perturbative and non-perturbative regime can be obtained from the results of 
appendix~\ref{sec:appendix_explicit_solutions}. Indeed, from eq.~\eqref{eq:dif_regimes} one infers that the boundary lies at
\begin{equation}
\label{bound} 
\langle A + \bar{A}\rangle = \left(-\frac{27}{2}p \hat{T} \lvert W_0 \lvert^2 \right)^{1/3}\ .
\end{equation}
We see that $\lvert \hat{T} \lvert \,\lvert W_0 \lvert^2$ has to be sufficiently large for some given $p$ to ensure that the minimum 
in eqs.~\eqref{eq:minimum_toy_model} still lies within the perturbative regime. For example, for $p=1$
one needs $\lvert \hat{T} \lvert \lvert W_0 \lvert^2 \,\gtrsim 10^{-3}$.
%

The existence of the minima in eq.~\eqref{eq:minimum_toy_model} are of particular interest in string theory, where the K\"ahler potential in
\eqref{eq:1dtoymodel} for $p=1$ typically describes the geometry of the dilaton.
For example in Calabi-Yau compactifications of the heterotic string the perturbative superpotential does not depend on the dilaton 
and background fluxes can generate a superpotential $W_0$, which is sufficiently big to ensure perturbativity. Of course a proper discussion of the dilaton in such scenarios lies outside the scope of this paper. We leave this to future research.
\section{Consequences for Moduli Stabilization in Type IIB}
\label{sec:LVS}
In this section we consider 
type IIB Calabi-Yau orientifold compactifications with background fluxes and 
the dynamics of the respective four-dimensional 
$\mathcal{N}=1$ K\"ahler moduli sector.
At lowest order in the effective action appropriate fluxes can stabilize the 
dilaton and complex structure moduli supersymmetrically, but the K\"ahler moduli are flat 
directions described by a no-scale model. 
The leading order $(\alpha')^3$-corrections in the bosonic ten-dimensional action 
include specific contractions of four Riemann-tensors \cite{Antoniadis:1997eg, Antoniadis:2003sw}. It was shown that 
these terms induce a correction to the K\"{a}hler potential of the K\"{a}hler moduli in the four-dimensional theory, that lifts the no-scale property \cite{Becker:2002nn}. 
Furthermore the K\"ahler potential can receive certain string-loop corrections. 
These have been explicitly computed for toroidal orientifolds, such as $T^6/(\mathbb{Z}_2 \times \mathbb{Z}_2)$ in
\cite{Berg:2005ja} and for arbitrary Calabi-Yau threefolds their functional form has been inferred in \cite{Berg:2007wt}. 

Besides the $R^4$ term the action of the type IIB superstring in ten-dimensions receives several 
additional contributions at order $(\alpha')^3$ of which a subset accounts for the ordinary ``two-derivative`` 
scalar potential after compactification. However, we will see that certain \mbox{$(\alpha')^3$-corrections} in ten dimensions exist, 
which contribute to the scalar potential in four dimensions, but which cannot be described as corrections to the K\"ahler or superpotential. Instead 
they demand the introduction of off-shell higher-derivative operators such as eq.~\eqref{eq:lagrangian_superspace_hd} as supersymmetric completions in four dimensions. 
The respective ten-dimensional pieces, which yield the additional corrections to the scalar potential after compactification, are not fully known and 
so we focus on the four-derivative partner-terms, which can in principle be computed exactly. More precisely we determine those four-derivative terms which are induced by the $R^4$ correction.\footnote{There may exist additional four-derivative terms which involve factors of the flux superpotential and hence the overall volume. These have to be merged into off-shell operators with more than four superspace-derivatives. Thus, the respective correction to the scalar potential from such terms is subleading.} We then infer the corrections to the scalar potential from eq.~\eqref{V1}. 
This identification is unique and will be discussed in a forthcoming publication.
The detailed computation of the four-derivative terms of the four-dimensional theory can be found in appendix~\ref{app:tensoralphaprime}. In this section we 
present the action in ten-dimensions and illustrate the influence of the individual terms on the theory in four dimensions. Afterwards we will display the resulting 
potential, which emerges from the results of appendix~\ref{app:tensoralphaprime} and study the possible implications for moduli stabilization these novel corrections might bring.
%
\subsection{Type IIB Action and Perturbative Corrections}\label{alphaprimeexp}
%
The low energy effective action of type IIB receives perturbative corrections in $\alpha'$ as well as in $g_s$. The leading order corrections to the action of the bulk fields arise at order $(\alpha')^3$ and consist of several eight-derivative terms. More specifically, the bosonic action takes the form
\begin{equation}\label{actionIIB10d}
 S_{IIB} = S_{b,0} + (\alpha')^3 S_{b,3} + \dots \ ,
\end{equation}
where $S_{b,0}$ denotes the tree level bosonic action of the bulk fields in the string-frame 
\begin{equation}\label{sb0}
 S_{b,0} = - \frac{1}{\kappa_{10}^2} \int \d^{10} x \sqrt{-g} \mathrm{e}^{-2\phi} \left(R + 4(\partial \phi)^2 - \frac{1}{2 \cdot 3!} H_3^2 \right) + S_{R} + S_{cs} \ .
\end{equation}
Eq.~\eqref{sb0} contains the ordinary kinetic terms for the bosonic fields of the type IIB superstring as well as the Chern-Simons term. Here we displayed explicitly the NS-NS sector which includes the metric $g$, the ten-dimensional dilaton $\phi$ and a two-form with field strength $H_3$. In eq.~\eqref{actionIIB10d} we neglected terms associated with localised sources, such as D3/D7 branes or O3/O7 orientifold planes. As the D3-branes are spacetime-filling they do not contribute to the scalar potential. Wrapped D7-branes on the other hand contribute and the leading order $(\alpha')^2$-corrections to their action are relevant and were discussed in \cite{Giddings:2001yu}. These corrections induce effective D3-brane charge and tension. Higher order $\alpha'$-corrections to the action of the localised sources can be ignored here \cite{Conlon:2005ki}. Recently additional $(\alpha')^2$-corrections to the K\"ahler potential for the K\"ahler moduli were inferred from F-theory \cite{Grimm:2013gma, Grimm:2013bha}. These corrections are related to a redundancy in the underlying M-theory description \cite{Junghans:2014zla} and can be absorbed via field-redefinitions \cite{Grimm:2013gma, Grimm:2013bha}.

The term $S_{b,3}$ in eq.~\eqref{actionIIB10d} contains the leading order, eight-derivative $\alpha'$-corrections to the action of the bulk fields. The full explicit structure of $S_{b,3}$ is unknown. Nevertheless one can infer their general form to be schematically \cite{Conlon:2005ki}
\begin{equation}\begin{aligned}\label{actionIIB}
S_{b,3} \sim \frac{1}{\kappa_{10}^2} \int& \d^{10} x \sqrt{-g} \Bigl[R^4 + R^3(G_3 G_3 + G_3 \bar{G}_3 + \bar{G}_3 \bar{G}_3  + F_5^2 + (\nabla \tau)^2) \\
 &+ R^2(G_3^4 + G_3^2 \bar{G}_3^2 + \dots + (\nabla G_3)^2 + (\nabla F_5)^2 + \dots ) \\
 &+ R(G_3^6 + \dots + G_3^2(\nabla G_3)^2 + \dots ) + G_3^8 + \dots \Bigr] \ .
\end{aligned}\end{equation}
Here $G_3$ is given by
\begin{equation}\label{G3}
 G_3 = F_3 - \tau H_3 \ ,
\end{equation}
where $F_3$ denotes the field strength of the RR two-form and $\tau$ is the axiodilaton, cf.~eq.~\eqref{N=1coords}. Moreover, $R$ schematically denotes the Riemann tensor and $\nabla G_3$ the covariant derivative (defined with respect to the metric $g$). Besides $G_3$ and $g$ the bosonic sector includes the axiodilaton $\tau$ as well as the self-dual five form field strength $F_5$. All indices within the terms in eq.~\eqref{actionIIB} are suppressed. Note that expressions with a single factor of $G_3$ or $F_5$ are forbidden. The precise structure of some of the contributions in eq.~\eqref{actionIIB} is explicitly known. Notably this is the case for the $R^4$ term to which we turn in a moment, but also all remaining quartic terms have been determined \cite{Kehagias:1997cq, Policastro:2006vt}. Furthermore couplings of the type $R^3 H_{3}^2$ and $R^2 H_{3}^4$ are required to ensure supersymmetry \cite{Liu:2013dna}. These terms imply the existence of the $R^3 G_3^2$, $R^3 G_3 \bar{G}_3$, $R^3 \bar{G}_3 \bar{G}_3$, $R^2G_3^2 \bar{G}_3^2$ and further contributions in eq.~\eqref{actionIIB}.

Note that the $R^4$ contribution in eq.~\eqref{actionIIB} is known exactly and has been determined in \cite{Antoniadis:1997eg}.\footnote{Note that also loop corrections contribute $R^4$-type terms which have been computed for instance in \cite{Green:1997as}. The tensor structure of these corrections is precisely the same as the tree-level term as required by \mbox{supersymmetry \cite{Antoniadis:1997eg}.}} This particular sum of contractions of four Riemann tensors is usually denoted as\footnote{At tree level this contraction is present for both IIA and IIB and all factors coincide.}
\begin{equation}
\label{J0}
 J_0 = t_8 t_8 R^4 + \tfrac{1}{8} \epsilon_{10} \epsilon_{10} R^4 \ .
\end{equation}
For the specific contractions in eq.~\eqref{J0} we refer to \cite{Antoniadis:1997eg}.

Eq.~\eqref{actionIIB} implies that contributions to the scalar potential in four dimensions exist, which involve four powers of the three-form flux $G_3$.\footnote{Note that in the situation with localised sources and background fluxes turned on we expect these contributions to be present. On the other hand, in the context of $\mathcal{N}=2$ compactifications these corrections will be absent as no scalar potential for the moduli is generated. Indeed the corrections to the potential that will be computed in this section vanish when turning off fluxes.}  
These terms cannot be captured via corrections to the K\"ahler or superpotential. Since the explicit form of the quartic terms in $G_3$ in ten dimensions are unknown, one can in principle not compute the respective correction to the potential directly. However, it turns out that the proper off-shell completion of this correction is given by the higher-derivative operator in eq.~\eqref{eq:lagrangian_superspace_hd}, which induces also four-derivative terms. These four-derivative terms are generated purely through $J_0$, which is known exactly and, thus, the structure of the correction to the potential can be inferred from eq.~\eqref{sugraL}. The explicit computation of the four-derivative term is performed in appendix~\ref{app:tensoralphaprime}. Here we will only summarize the main steps. For simplicity only a single K\"ahler class deformation is turned on. However, we expect the inferred form of the correction to the scalar potential $V_{(1)}$ in eq.~\eqref{V1twoB} to hold also in the case of arbitrarily many K\"ahler moduli (we explain this at the end of appendix~\ref{app:tensoralphaprime}), as long as the superpotential does not receive non-perturbative corrections. Moreover, due to the presence of background fluxes the background metric has to involve a warp factor. Here we are interested in the behaviour of the potential at large volume and, therefore, we work in a weak-warping approximation in which we neglect all warping effects.

We set the four-dimensional piece of the metric to a Minkowski-form. Neglecting the warping the ten-dimensional metric in the string-frame then reads
\begin{equation}\label{10dmetricalpha}
 \d s_{(10)}^2 =  \eta_{\mu \nu} \d x^{\mu} \d x^{\nu} + \hat{\mathcal{V}}^{1/3} (x) \d s_{(6)}^2\ , 
\end{equation}
where $\hat{\mathcal{V}}$ describes the K\"ahler type deformation of the (string-frame) background metric of the Calabi-Yau threefold denoted by $\d s_{(6)}^2$. The next step then involves the computation of the components of the Riemann tensor and finally we determine the four-derivative terms for $\hat{\mathcal{V}}$, which emerge from $J_0$. Afterwards it is necessary to express the result in terms of the appropriate $\mathcal{N}=1$ variables and match to the four-derivative term inside the Lagrangian in eq.~\eqref{sugraL} to determine the form of $T_{ij \bar{k} \bar{l}}$. We will present the result in the next section. First it is necessary to establish the notation of the $\mathcal{N}=1$ theory and discuss the 
known contributions to the scalar potential.
\subsection{Structure of Scalar Potential}
Let us now proceed to discuss the general structure of the scalar potential 
including the known leading order $\alpha'$- and string-loop corrections to the K\"ahler 
potential as well as the new, 'higher-derivative' $\alpha'$-corrections.
Concretely we consider the K\"ahler and superpotential that arises
after integrating out the complex structure moduli and 
the dilaton
\begin{equation}\begin{aligned}
\label{eq:kahler_moduli_sugra}
W = W_0 = \Bigl\langle \int_{M_3} G_3 \wedge \Omega \Bigr\rangle \ ,\qquad
K = - 2 \,\mathrm{ln} \left(\mathcal{V} + \frac{\hat{\xi}}{2} \right) + \delta K^{KK}_{(g_s)} + \delta K^{W}_{(g_s)}\ . 
\end{aligned}\end{equation}
Here $W_0$ denotes the flux superpotential, which is the Gukow-Vafa-Witten superpotential evaluated at the supersymmetric minimum of the complex structure moduli and the dilaton. Furthermore, $G_3$ is given in eq.~\eqref{G3} and $\Omega$ is the $(3,0)$ form of the Calabi-Yau. The total (Einstein-frame) volume modulus
$\mathcal{V}$ can be expressed in terms of the (completely symmetric) triple intersection numbers $k_{ijk}$ of the Calabi-Yau $M_3$ as well as the 2-cycle volumes $t^i$ as follows
\begin{equation}\label{eq:volume_of_cy}
\mathcal{V} = \frac{1}{6} k_{ijk} t^i t^j t^k \ .
\end{equation}
The 4-cycle volumes $\tau_i$, that constitute the imaginary components
of the K\"{a}hler moduli $T^i$, are derived via 
\begin{equation}\label{deffourcycle}
 \tau_i = \frac{\partial \mathcal{V}}{\partial t^i} = \frac{1}{2} k_{ijk} t^j t^k \ .
\end{equation}
From these definitions one infers
\begin{equation}\label{Veqttau}
 \mathcal{V} = \frac{1}{3} \tau_i t^i \ .
\end{equation} 
Furthermore $\hat{\xi}$ parametrizes the leading $\alpha'$-corrections to the K\"{a}hler potential and is given by
\begin{equation}\label{xi}
 \hat{\xi} = \xi g_s^{-3/2} = - \frac{(\alpha')^3 \zeta(3) \chi(M_3)}{2 (2\pi)^3 g_s^{3/2}} \ ,
\end{equation}
where $\chi(M_3)=2(h^{1,1}-h^{2,1})$ is the Euler characteristic of
$M_3$, $g_s$ denotes the string-coupling and  the Hodge numbers
$h^{1,1},h^{2,1}$
count the number of K\"ahler and complex structure moduli.
The corrections $\delta K^{KK}_{(g_s)}$ and $\delta K^{W}_{(g_s)}$ in eq.~\eqref{eq:kahler_moduli_sugra} 
denote the leading order string-loop corrections. Their general form for arbitrary Calabi-Yau threefolds has been argued to be~\cite{Berg:2007wt} 
\begin{equation}\begin{aligned}
\label{eq:string_loop_corrections}
 \delta K^{KK}_{(g_s)} &\sim g_s \sum_{i=1}^{h^{1,1}} 
\frac{C_i(a_{ij}t^j)}{\mathcal{V}} \ , \qquad
 \delta K^{W}_{(g_s)} \sim \sum_{i=1}^{h^{1,1}} \frac{D_i (a_{ij}t^j)^{-1}}{\mathcal{V}} \ .
\end{aligned}\end{equation}
The first term is interpreted as coming from exchange of closed strings carrying Kaluza-Klein momentum, while the latter is coming from the exchange of winding strings. The coefficients $C_i$ and $D_i$ are expected to be functions of the complex structure moduli and the dilaton. However, since we assume the latter have already been stabilized, we treat $C_i, D_i$ as constants. The matrix $a_{ij}$ consists of combinatorial constants.

The scalar potential derived from eq.~\eqref{eq:kahler_moduli_sugra} including the higher-derivative term $\mathcal{L}_{(1)}$ in \eqref{eq:lagrangian_superspace_hd} can be split up as follows
\begin{equation}\begin{aligned}
\label{totalpot}
  V = V_{(0)} + V_{(1)} = V_{(\alpha')} + V_{(g_s)} + V_{(1)}  \ .
\end{aligned}\end{equation}
The first term describes the scalar potential obtained from the K\"ahler potential in eq.~\eqref{eq:kahler_moduli_sugra} without string-loop corrections. It reads \cite{Balasubramanian:2004uy} 
\begin{equation}
 \label{Valphaprime}
  V_{(\alpha')} = \mathrm{e}^K 3 \,\hat{\xi} \lvert W_0 \lvert^2\, \frac{\hat{\xi}^2+7\hat{\xi}\mathcal{V} + \mathcal{V}^2}{(\mathcal{V}-\hat{\xi})(2\mathcal{V}+\hat{\xi})^2} \ ,
\end{equation}
and has a runaway behaviour at large $\mathcal{V}$. Expanding around large volume yields  
\begin{equation}
\label{ValphaprimeLV}
 V_{(\alpha')} =  \frac{3 \hat{\xi} \lvert W_0 \lvert^2}{4 \mathcal{V}^3} + \mathcal{O}((\alpha')^6) \ .
\end{equation}
When expanding the string-loop contribution to the potential, one obtains the following terms at leading order \cite{Cicoli:2007xp}
\begin{equation}
 V_{(g_s)} = \sum_i \frac{\lvert W_0 \lvert^2}{\mathcal{V}^2} \left[ g_s^2 C_i^2 K_{(0),ii} - 2 \partial_{\tau_i}(\delta K^{W}_{(g_s)}) \right] \ ,
\end{equation}
where $K_{(0)} = -2 \mathrm{ln}(\mathcal{V})$.

Let us now turn to the higher-derivative operator. Inserting eq.~\eqref{eq:kahler_moduli_sugra} into eq.~\eqref{V1} the higher-derivative 
contribution generally has the form
\begin{equation}\label{V1gen}
 V_{(1)} = -  \mathrm{e}^{2K} T^{\bar{i}\bar{j} kl} K_{,\bar{i}} K_{,\bar{j}} K_{,k} K_{,l} \lvert W_0 \lvert^4 \ .
\end{equation}
The result of appendix~\ref{app:tensoralphaprime}
are four-derivative terms for the four-cycle volumes, which when matched to eq.~\eqref{sugraL} yield 
the following coupling tensor 
\begin{equation}
\label{resultT}
 T_{ij \bar{k} \bar{l}} = \hat\lambda_0  (\Pi_m \, t^m) K_{(0),i} K_{(0),j} K_{(0),\bar{k}} K_{(0),\bar{l}} \ ,
\end{equation}
where we introduced $\hat{\lambda}_0 = (\alpha')^3 g_s^{-3/2} \lambda$ with $\lambda$ being
a universal combinatorial number that is not computed at this stage. A direct 4D reduction of the partially unknown 10D terms with four powers in the fluxes and their derivatives which contribute to the scalar potential of ${\cal L}_{(1)}$ would determine the sign of $\lambda$.  Thus, at this point we treat it as an unknown real number. 
Note furthermore that \mbox{$\hat{\lambda}_0 \sim (\alpha')^3 g_s^{-3/2}$} includes
precisely the same expansion parameter that shows up in the correction
to the K\"ahler potential
via $\hat{\xi}$ given in \eqref{xi}. This is expected since both corrections 
originate from the same term in ten dimensions.
Let us mention again that the result in eq.~\eqref{resultT} holds only for $h^{1,1}=1$. However, 
we expect the respective correction to the potential to be correct in general as we discuss at the 
end of appendix~\ref{app:tensoralphaprime} and in appendix~\ref{app:geometrytensor}. 
The numbers $\Pi_i$ encode the topological information of the second Chern class $c_2$ of~$M_3$. 
Specifically let us choose a basis $\hat{D}_i$ of harmonic $(1,1)$-forms, such that 
the K\"ahler form is expressed as 
\begin{equation}\label{Kahlerform}
J = \sum_{i=1}^{h^{1,1}} \hat{D}_i t^i \ .
\end{equation}
Then we have that
\begin{equation}
 \Pi_i = \int_{M_3} c_2 \wedge \hat{D}_i \ ,\qquad \Pi_i t^i =
 \int_{M_3} c_2 \wedge J\ .
\end{equation}
In appendix~\ref{app:tensoralphaprime} we discuss this term further. 
In particular we have $\int_{M_3} c_2 \wedge J > 0$ unless $M_3$ is a torus $T^6$.
The variables in which $J$ takes the form of eq.~\eqref{Kahlerform} span the K\"ahler cone and, thus, 
we have $t_i\geq 0$ independently for all two-cycle volumes, see e.g. \cite{Denef:2008wq}. 
Accordingly, in order to ensure $\int_{M_3} c_2 \wedge J > 0$, we must have that $\Pi_i \geq 0$ in this basis. 
In section~\ref{MInimum} we compute these topological numbers for an explicit example.

Inserting \eqref{resultT} into eq.~\eqref{V1gen} we can read off the correction to the potential
\begin{equation}
\label{V1twoB}
 V_{(1)} = - \hat{\lambda} \frac{\lvert W_0 \lvert^4}{\mathcal{V}^4} (\Pi_i \, t^i) \ ,
\end{equation}
where we abbreviate $\hat{\lambda} = 3^4 \hat{\lambda}_0$. To understand the volume-behaviour of the individual terms in the potential \eqref{totalpot} in the large volume limit we set $h^{1,1}=1$. In this case we obtain
\begin{equation}
\label{eq:pot_h11=1}
 \frac{V}{\lvert W_0 \lvert^2} \sim \hat{\xi}  \mathcal{V}^{-3} + g_s^2 C_1^2 \mathcal{V}^{-10/3} -\hat{\lambda} \lvert W_0 \lvert^2 \Pi_1  \mathcal{V}^{-11/3} + D_1 \mathcal{V}^{-4} \ ,
\end{equation}
where we ignored numerical factors. The scaling behaviour of the higher-derivative contribution lies precisely in between the two string-loop contributions. It differs by a factor of $\lvert W_0 \lvert^2$ and by powers of the string coupling $g_s^{-3/2}$ and $g_s^{-7/2}$ with respect to the string-loop corrections. 

Before studying the implication of $V_{(1)}$ for K\"ahler moduli stabilization, 
we should pause for a moment to present a better understanding of the individual 
pieces of the ten-dimensional action, which was displayed in eq.~\eqref{actionIIB10d} and eq.~\eqref{actionIIB}. 
As already mentioned the remaining terms in eq.~\eqref{sugraL}, such as the corrections to
the scalar potential and to the ordinary kinetic term are related to different,
partially unknown terms in the ten dimensional action, which are connected by supersymmetry.
Furthermore we wish to analyze the relevance of further higher-order corrections in the large-volume expansion.

Let us see how the individual terms in eq.~\eqref{actionIIB} contribute to the four-dimensional action after compactification.~$J_0$ generates derivative terms for the K\"ahler moduli, but does not contribute to the potential if warping is neglected. This is due to the fact, that after turning off fluxes this term is still present. However, in the respective $\mathcal{N} = 2$ theory, no potential can be generated for the moduli as all $\alpha'$-corrections merely renormalise the definition of the tree-level moduli. Thus, $J_0$ induces derivative-corrections, such as the $\hat{\xi}$-contribution to the K\"ahler potential in eq.~\eqref{eq:kahler_moduli_sugra}, the four-derivative terms that we computed in appendix~\ref{app:tensoralphaprime} and further six- and eight-derivative terms. The $\hat{\xi}$-corrections imply the existence of the potential in eq.~\eqref{ValphaprimeLV} in the $\mathcal{N}=1$ theory. It was noted in \cite{Becker:2002nn} that after transforming into the Einstein-frame in the four-dimensional action the $H_3^2$ term in eq.~\eqref{sb0} indeed produces the correct functional form of eq.~\eqref{ValphaprimeLV}. However, to obtain the correct prefactors it was concluded that necessarily also $R^3 G_3^2$ terms have to be present. Indeed, we have 
\begin{equation}
 V_{(\alpha')} \sim \underbrace{\chi(M_3)}_{R^3}  \underbrace{\lvert W_0 \lvert^2}_{G_3 \bar{G}_3} \ ,
\end{equation}
where we used that $\chi(M_3) \sim \int \d^6 y \sqrt{g} Q$ with the six-dimensional Euler integrand $Q$ being a contraction of three Riemann tensors. The corrections of the type $R^2 (\nabla G_3)^2$ also contribute to $V_{(\alpha')}$.

Next let discuss the additional terms in the Lagrangian which accompany the four-derivative terms in eq.~\eqref{sugraL}. The non-K\"ahler correction to the two-derivative term in eq.~\eqref{deltaGSUGRA} is induced by the terms of the type $R^3 G_3 \bar{G}_3$. These corrections have the form
\begin{equation}\label{nonkahlercorrIIB}
 \delta G_{i \bar{j}} \,\partial_{\mu} T^i \partial^{\mu} \bar{T}^{\bar{j}} \sim \frac{\hat{\lambda}}{\mathcal{V}^2}\underbrace{\lvert W_0 \lvert^2}_{\sim G_3 \bar{G}_3}  \underbrace{\Pi_m t^m}_{\sim R^2}  \underbrace{K_{(0),i} K_{(0),\bar{j}} \partial_{\mu} T^i \partial^{\mu} \bar{T}^{\bar{j}}}_{\sim R} \ ,
\end{equation}
where we used eq.~\eqref{c2wedgeJ}. Furthermore the terms with two Riemann tensors in eq.~\eqref{actionIIB} generate $V_{(1)}$ in eq.~\eqref{V1twoB}, since 
\begin{equation}
 V_{(1)} \sim \frac{\hat{\lambda}}{\mathcal{V}^4} \underbrace{\lvert W_0 \lvert^4}_{\sim G_3^2 \bar{G}_3^2} \underbrace{(\Pi_i \, t^i)}_{\sim R^2} \ .
\end{equation}
Furthermore $V_{(1)}$ can be induced by terms of the type $R G_3^2(\nabla G_3)^2$.\footnote{Note that in \cite{Conlon:2005ki} a naive estimate for the volume dependence of the potential induced by the $R^2 G_3^4$ terms was found to be $\mathcal{V}^{-11/3}$. This is in agreement with eq.~\eqref{V1twoB}.}

Let us now make a few remarks regarding the terms we did not discuss so far. To begin with there exist corrections with additional derivatives of the dilaton. These terms do not contribute to the scalar potential, but are important for the consistency of the equations of motion. More precisely the presence of the $R^4$ terms demands the addition of terms of the type $R^3 (\nabla \tau)^2$ \cite{Becker:2002nn}. Furthermore we have terms involving the self-dual five-form $F_5$. In compactifications with imaginary self-dual fluxes warping effects generate a flux for the five-form \cite{Giddings:2001yu}. Since we ignore the warp factor here, we will not discuss this term further. However, in principle warping-induced corrections to the scalar potential are relevant, since naive dimensional arguments suggest that these contribute at $\mathcal{O}(\mathcal{V}^{-11/3})$ \cite{Conlon:2005ki}. A proper accounting of such effects is outside the scope of this paper and will be left to future investigations.
Moreover we have terms of the type $R G_3^6$ and $G_3^8$ in eq.~\eqref{actionIIB}. Dimensional analysis yields that the contributions to the scalar potential coming from both terms are suppressed by additional powers of $\mathcal{V}^{-2/3}$ and $\mathcal{V}^{-4/3}$ with respect to $V_{(1)}$ \cite{Conlon:2005ki}. Furthermore naively one finds that a reduction of $R G_3^6$ yields a factor of $c_1(M_3)$, which vanishes for a Calabi-Yau orientifold compactification at order $(\alpha')^3$, see also appendix~\ref{app:tensoralphaprime}.

The potential in eq.~\eqref{Valphaprime} also induces subleading terms at the level of $(\alpha')^6$, which scale as $\mathcal{V}^{-4}$. Besides the fact that their volume-dependence is slightly suppressed compared to $V_{(1)}$, they involve a factor $\hat{\xi}^2$, which is rather small for CY threefolds with small Euler number and moderate $g_s$-values.

To conclude this section let us make a remark regarding the expansion in higher-derivatives of the action in eq.~\eqref{actionIIB}. The expansion in $\alpha'$ in ten dimensions is indeed an expansion in higher-derivatives. However, when compactifying the $R^4$ term in eq.~\eqref{actionIIB}, one obtains two, four, six and eight-derivative terms for the volume modulus, cf.~appendix~\ref{app:tensoralphaprime}. Thus, in the four dimensional theory the $\alpha'$-expansion is still roughly controlling the expansion in higher-derivatives, but several higher-derivative terms might appear at the same order in $\alpha'$. This implies that the coupling tensor in eq.~\eqref{resultT} cannot control all higher-derivatives, but possibly only a subclass. Moreover, let us briefly revisit the general discussion in sec.~\ref{onechiral}, as we now have an example with an explicit expansion parameter given by $\alpha'$. Recall that we identified the analytic branch as the unique physical theory. Since $T_{ij\bar{k} \bar{l}} \sim (\alpha')^3$ we find evidence for this once more. In particular the non-analytic branches would require the presence of terms in ten-dimensions, which are $\mathcal{O}(\alpha'^{-3})$. Furthermore note that it would not be meaningful to discuss the corrections at order $\mathcal{O}(T^2)$, as we would have to include  ten-dimensional terms of order $(\alpha')^6$ into the analysis.
\subsection{Stabilization of the Volume for $h^{1,1}=1$}
Using the leading order $\alpha'$-correction to the K\"{a}hler potential
accompanied by non-perturbative corrections to the superpotential there exist scenarios, where all K\"{a}hler moduli can be frozen \cite{Kachru:2003aw, Balasubramanian:2004uy, Balasubramanian:2005zx, Louis:2012nb}. 
Later works incorporated also string-loop-corrections in the K\"ahler potential 
into the analysis \cite{Berg:2005ja, Berg:2005yu, Berg:2007wt, Cicoli:2007xp, Cicoli:2008va}. 
In all of these scenarios the non-perturbative 
superpotential is necessary for the stability of the overall volume.
Attempts to stabilize the volume modulus without the non-perturbative superpotential 
including string-loop corrections were made in \cite{Berg:2005yu}, but 
a significant amount of fine-tuning of the complex-structure moduli was required. 
In addition the structure of the string-loop corrections is very model-dependent 
and a case-by-case study is necessary.

In the following we will entertain the possibility that 
the overall volume and all four-cycle volumes are stabilized purely 
by $\alpha'$-corrections instead of the 
non-perturbative corrections to the superpotential. The leading order 
$(\alpha')^3$-corrections are partially captured by the higher-derivative corrections 
together with the known corrections to the K\"ahler potential.
It is instructive to discuss a stabilization first 
in the simple case of $h^{1,1}=1$. We will generalize the analysis to 
an arbitrary number of four-cycles in the next section.
In the following we neglect string-loop corrections to the scalar potential. 
Since these are suppressed by powers $g_s^{3/2}$ and $g_s^{7/2}$, respectively, 
relative to both $(\alpha')^3$-contributions, a moderate tuning of $g_s < 1$ should 
suffice to parametrically suppress them. Moreover, note that
from the discussion in refs.\ \cite{Berg:2005ja, Berg:2005yu} it is also
expected that the coefficients $C_i$ and $D_i$ in eq.~\eqref{eq:string_loop_corrections} are small. 
Indeed, in the explicit computations they are suppressed by loop factors of $1/(128 \pi^4)$ 
and thus small, unless the complex structure moduli are frozen at large values.

The potential in eq.~\eqref{eq:pot_h11=1} is then minimized at
\begin{equation}
\label{eq:minimum_volume}
 \langle \mathcal{V} \rangle \sim \left( \frac{\hat{\lambda} \lvert W_0 \lvert^2 \Pi_1 }{\hat{\xi}} \right)^{3/2}
\end{equation}
under the assumption that $\hat{\lambda} < 0$ and $\hat{\xi}<0$. The latter requirement is fulfilled for any Calabi-Yau with $\chi(M_3) > 0$ or in other words $h^{1,1}(M_3) > h^{2,1}(M_3)$. Note that background fluxes require $h^{1,1}(M_3) > h^{2,1}(M_3) \geq 1$. Hence, we need at least two K\"{a}hler moduli to satisfy $\chi(M_3) > 0$. Thus, the above analysis is not realistic. However, based on this simple example one would naively expect that in the case of $h^{1,1}>1$ one finds 
a stabilized volume only if $\chi(M_3) > 0$ and $\lambda < 0$. 
This is indeed confirmed in the next section.

Supersymmetry is broken in the vacuum given in eq.~\eqref{eq:minimum_volume} which can be seen as follows. From sec.~\ref{sec:susy_flat_dir} we know that supersymmetry is broken, if it is broken at two-derivative level. Suppose supersymmetry was unbroken, then one could derive the minimum from eq.~\eqref{sugramin}. However, necessarily all such points would be $\hat{\lambda}$-independent, which is not satisfied for our minimum. Thus, supersymmetry is indeed broken in the vacuum in eq.~\eqref{eq:minimum_volume}. Furthermore computation shows that it is an AdS vacuum with a value of the cosmological constant given by
\begin{equation}
 \langle V \rangle \sim \frac{\hat{\xi} \lvert W_0 \lvert^2}{\langle \mathcal{V} \rangle^3} < 0
\end{equation}
In the next section we generalize to the case $h^{1,1}>1$ and prove the existence of a general minimum. Finally, we note that this minimum does not arise by balancing two terms in the same expansion at different order. Instead, both terms are precisely of the same order in $\alpha'$ and $g_s$. Moreover, in the four-dimensional theory we formally have an expansion in the coupling tensor, which controls the higher-derivative corrections, as well as in $\hat{\xi}$, which controls $V_{(\alpha')}$ in eq.~\eqref{Valphaprime}. From this point of view, in the minimization we are comparing leading order terms, which are associated with different expansions.
\subsection{Existence of Model-Independent Minimum}\label{MInimum}

Neglecting string-loop corrections
and taking the large-volume limit the potential given in \eqref{totalpot} reads 
 \begin{equation}
 \label{eq:potLV}
  V = \frac{3\hat{\xi} \lvert W_0 \lvert^2}{4 \mathcal{V}^3}-\hat{\lambda}\lvert W_0 \lvert^4\, \frac{\Pi_i t^i}{\mathcal{V}^4} \ .
 \end{equation}
For $\hat\lambda<0$ we will now show that $V$ 
has a non-supersymmetric AdS minimum for any orientifolded Calabi-Yau threefold with 
$\chi(M_3) >0$ where \emph{all} four-cycles are fixed as
\begin{equation}\label{taumin}
 \langle \tau_i \rangle = \mathcal{C} \,\Pi_i \ ,\qquad \textrm{with} \qquad \mathcal{C} = \frac{44 \hat{\lambda} \lvert W_0 \lvert^2 }{9 \hat{\xi}}\ .
\end{equation}
The volume in this minimum is given by
\begin{equation}\label{Vmin}
 \langle \mathcal{V} \rangle = \tfrac{1}{3} \mathcal{C} \, \Pi_k \langle t^k \rangle =  \frac{44}{27} \left\langle \int c_2 \wedge J \right\rangle\, \frac{\hat{\lambda} \lvert W_0 \lvert^2}{\hat{\xi}}\ \sim\ \Pi_k \langle t_{0}^k \rangle \left( \frac{\hat{\lambda} \lvert W_0 \lvert^2}{\hat{\xi}} \right)^{3/2} \ ,
\end{equation}
where $\langle t_{0}^i \rangle$ do not depend on $\mathcal{C}$, but are implicit functions of the $\Pi_i$. Moreover, positivity of the four-cycles requires that $\Pi_i>0$ for all $i=1, \dots, h^{1,1}$. As we already mentioned when choosing the correct K\"ahler cone variables one has $\Pi_i \geq 0$, so we have to require that $\Pi_i \neq 0$.

In order to prove the existence of this minimum
it is sufficient to show that the potential in eq.~\eqref{eq:potLV} is minimal as a function of the two-cycle volumes $t^i$ as it is then also minimal in terms of the four-cycle volumes $\tau_i$. The first derivatives of eq.~\eqref{eq:potLV} read
\begin{equation}
 \frac{\partial V}{\partial t^i} = \frac{\lvert W_0 \lvert^2}{\mathcal{V}^5} \left[-\frac{3}{4} \hat{\xi} \tau_i \left(t^i \tau_i \right) - \frac{1}{3} \hat{\lambda} \lvert W_0 \lvert^2 \Pi_i \left( t^j \tau_j \right) + 4\hat{\lambda}\lvert W_0 \lvert^2 \tau_i \left( \Pi_j t^j \right) \right] \ ,
\end{equation}
where we used eq.~\eqref{Veqttau}. Inserting the values of the four-cycle volumes given in eq.~\eqref{taumin} one finds that indeed $\langle \partial V / \partial t^i \rangle = 0$. From eq.~\eqref{Veqttau} we also obtain the first equality in eq.~\eqref{Vmin}. To determine the overall dependence of $\langle \mathcal{V} \rangle$ on $\mathcal{C}$, note that the two-cycles are implicitly defined via eq.~\eqref{deffourcycle}, which at the extremal point is given by
\begin{equation}
\label{eq:tt=Pi}
 k_{ijk} \langle t^j \rangle \langle t^k \rangle = 2 \mathcal{C} \Pi_i \ .
\end{equation}
This implies $\langle t^i \rangle = \sqrt{\mathcal{C}} \langle t_{0}^i \rangle$, where $t_{0}^i$ do not depend on $\mathcal{C}$. With this we obtain the scaling of the volume with respect to $\lvert W_0 \lvert, \hat{\xi}$ and $\hat{\lambda}$ in eq.~\eqref{Vmin}.

It remains to analyse the matrix of second derivatives. In general it reads
\begin{equation}\begin{aligned}
  \frac{\partial^2 V}{\partial t^i \partial t^j} =  \frac{\lvert W_0 \lvert^2}{\mathcal{V}^6} \Bigl[&9 \hat{\xi}\mathcal{V}\tau_i \tau_j + 4 \hat{\lambda} \lvert W_0 \lvert^2 \mathcal{V}\left( \tau_i \Pi_j + \Pi_i \tau_j \right) - 20\hat{\lambda} \lvert W_0 \lvert^2 (\Pi_k t^k) \tau_i \tau_j \\
& + \frac{\partial \tau_j}{\partial t^i}\left(4\hat{\lambda}  \lvert W_0 \lvert^2 \mathcal{V}(\Pi_k t^k) - \frac{9}{4}\hat{\xi}\mathcal{V}^2 \right)\Bigr] \ .
\end{aligned}\end{equation}
Making use of eq.~\eqref{Veqttau} we find that at the extremal point this simplifies to
\begin{equation}\begin{aligned}\label{secdermin}
 \left\langle  \frac{\partial^2 V}{\partial t^i \partial t^j}  \right\rangle = a \Pi_i \Pi_j + b k_{ijk}  \langle t^k \rangle \ , \quad \textrm{where} \qquad a = -\frac{8\hat{\lambda} \lvert W_0 \lvert^4 \mathcal{C} }{\langle \mathcal{V} \rangle^5} \ , \qquad b= \frac{9}{44}\frac{\hat{\xi}\lvert W_0 \lvert^2}{\langle \mathcal{V} \rangle^4} \ .
\end{aligned}\end{equation}
For $\lambda<0$ and $\chi(M_3) >0$ we see that $a>0$ and $b<0$. For any vector with components $x_i$ we have $(x_i \Pi_i) (x_j \Pi_j) \geq 0$ and so $a \Pi_i \Pi_j$ is a positive-semidefinite matrix. The matrix $k_{ijk} t^k$ was studied 
in \cite{Candelas:1990pi} and shown to have signature $(1,h^{1,1}-1)$. In other words there exists an orthogonal decomposition of the $h^{1,1}$-dimensional vector space into a one-dimensional subspace, on which $k_{ijk} t^k$ is positive definite and an $(h^{1,1}-1)$-dimensional complement on which it is negative definite. Here orthogonality is defined with respect to the inner product determined by $k_{ijk} t^k$. The one-dimensional subspace is spanned by the vector with components $t^i$, as the volume has to be positive.  Since we have $b<0$
the signature of $b k_{ijk}  \langle t^k \rangle$ reads $(h^{1,1}-1,1)$. On the $(h^{1,1}-1)$-dimensional subspace the sum $a \Pi_i \Pi_j + b k_{ijk}  \langle t^k \rangle$ must hence be positive-definite.
On the one-dimensional subspace we find
\begin{equation}
\langle t^i \rangle \left\langle  \frac{\partial^2 V}{\partial t^i \partial t^j}  \right\rangle \langle t^j \rangle = -\frac{22 \hat{\lambda} \lvert W_0 \lvert^4 \mathcal{C} }{3\langle \mathcal{V} \rangle^5} \left( \Pi_k  \langle t^k \rangle \right)^2 > 0 \ ,
\end{equation}
which shows that the matrix of second derivatives is also positive definite there. 

It remains to be shown, that the matrix \eqref{secdermin} is positive definite on the whole space. A generic non-zero vector with components $x^i$ can be decomposed as $x^i = \mu \langle t^i \rangle + x_{\perp}^i$, where $\mu \in \mathbb{R}$ 
and $x_{\perp}^i$ is the component of $x^i$ in the subspace orthogonal to the one-dimensional space spanned by $\langle t^i \rangle$. Since
\begin{equation}
\Pi_i \Pi_j \langle t^j \rangle \sim \Pi_i \sim k_{ijk}  \langle t^j \rangle \langle t^k \rangle
\end{equation}
we have the following orthogonality relations
\begin{equation}
x_{\perp}^i k_{ijk}  \langle t^j \rangle \langle t^k \rangle =  x_{\perp}^i \Pi_i \Pi_j \langle t^j \rangle = 0 \ .
\end{equation}
With this we find
\begin{equation}\begin{aligned}
 x^i \left\langle  \frac{\partial^2 V}{\partial t^i \partial t^j}  \right\rangle x^j = x_{\perp}^i \left( a \Pi_i \Pi_j + b k_{ijk}  \langle t^k \rangle \right) x_{\perp}^j + \mu^2 \langle t^i \rangle \left( a \Pi_i \Pi_j + b k_{ijk}  \langle t^k \rangle \right) \langle t^j \rangle > 0\ ,
\end{aligned}\end{equation}
since the matrix is positive on the respective subspaces. 
We conclude that the matrix in eq.~\eqref{secdermin} is positive definite.

In addition we have to establish that the locus specified in eq.~\eqref{taumin} is a minimum of the potential, which includes also the dilaton as well as the complex structure moduli. The answer can be easily obtained in the spirit of \cite{Balasubramanian:2005zx}. Indeed the potential including the dilaton and complex-structure moduli reads \cite{Becker:2002nn}
\begin{equation}\label{fullpot}
  V = \mathrm{e}^K (G^{a\bar{b}} D_a W D_{\bar{b}} \bar{W} + G^{\tau \bar{\tau}} D_{\tau} W D_{\bar{\tau}} \bar{W})+ \mathrm{e}^K \frac{\xi}{2 \mathcal{V}}(WD_{\bar{\tau}}  \bar{W} + \bar{W} D_{\tau} W) + V_{(\alpha')} +V_{(1)} \ ,
\end{equation}
where $a,b$ label complex structure moduli directions. $W$ denotes the the Gukov-Vafa-Witten superpotential. The first term in the above potential is positive definite and has a $\mathcal{V}^{-2}$ behaviour at large volume. At the extremal condition $D_a W=D_{\tau}W=0$, it vanishes identically and is positive around this value. Since it dominates over the subleading $\mathcal{O}(\mathcal{V}^{-3})$ and $\mathcal{O}(\mathcal{V}^{-11/3})$ terms coming from $V_{(\alpha')} +V_{(1)}$, eq.~\eqref{taumin} represents a minimum of the full potential. Of course also the dilaton and complex structure moduli will receive higher-derivative corrections, which contribute in $V_{(1)}$. However, these terms 
have a subleading volume-dependence compared to the first terms in eq.~\eqref{fullpot} and thus do not spoil the argument.

As in the preceding section, supersymmetry is broken in the minimum. Up to numerical factors the value of the potential in the minimum reads
\begin{equation}
 \langle V \rangle \sim \frac{\hat{\xi}}{\lvert W_0 \lvert^7} \left(\frac{\hat{\xi}}{\hat{\lambda}}\right)^{9/2} \ .
\end{equation}
We can estimate the gravitino mass from the ordinary two-derivative theory. It reads  
\begin{equation}\label{gravmass}
m_{3/2} \sim \mathrm{e}^{K/2} \lvert W_0 \lvert \, \sim  \frac{\lvert W_0 \lvert }{\mathcal{V}}  \sim  \frac{\hat\xi^{3/2}}{\hat\lambda^{3/2} \lvert W_0 \lvert^{2}  \Pi_i \langle t_0^i \rangle} \ .
\end{equation}
Note that the corrections $F_{(1)}$ contribute only subleading here. Let us compare the gravitino mass with the string scale and Kaluza-Klein scale \cite{Conlon:2005ki}
\begin{equation}
 m_s \sim \frac{1}{\sqrt{\mathcal{V}}} \ , \qquad m_{KK} \sim \frac{1}{\mathcal{V}^{2/3}} \ .
\end{equation}
Direct computation reveals that
\begin{equation}
 \frac{m_{3/2}}{m_s} \sim \frac{\hat\xi^{3/4}}{\hat\lambda^{3/4} \sqrt{\lvert W_0 \lvert  \Pi_i \langle t_0^i \rangle}} \ .
\end{equation}
Furthermore, from eq.~\eqref{eq:tt=Pi} we find that roughly $\langle t_0^i \rangle \sim \sqrt{\Pi_i}$. Let $\Pi$ denote a typical value for the topological numbers $\Pi_i$, then we can estimate
\begin{equation}
 \Pi_i \langle t_0^i \rangle \sim h^{1,1} \Pi^{3/2} \ .
\end{equation}
In the next section we show that $\Pi \sim {\mathcal O}(10\ldots 100)$. Furthermore, we can estimate the size of $\hat{\lambda}$ by the combinatorial part of $\hat{\xi}$. In other words we roughly expect that $\lvert \hat{\lambda} \lvert \sim \lvert \hat\xi / \chi(M_3) \lvert$. Altogether, the scale-quotients read\footnote{We thank Shanta de Alwis for helpful comments and discussions regarding this point.} 
\begin{equation}
 \frac{m_{3/2}}{m_s} \sim \frac{\chi(M_3)^{3/4}}{\sqrt{\lvert W_0 \lvert \, h^{1,1} \, \Pi}}\lesssim \mathcal{O}(10^{-1}) \ , \qquad \frac{m_{3/2}}{m_{KK}} \sim \frac{{\chi(M_3)}^{1/2}}{(h^{1,1})^{1/3} \sqrt{\Pi}} < 1 \ .
\end{equation}
To obtain more accurate expressions for $m_{3/2} / m_s$ and $m_{3/2} / m_{KK}$, it will be necessary to compute $\hat\lambda$ and study the minimum for explicit examples.

Let us finish this section with some remarks. Firstly let us stress again that the stabilization of the four-cycle volumes proposed here does not require any non-perturbative effects, but occurs purely from considering the leading order $(\alpha')^3$-corrections in the potential. Note furthermore that even though a Calabi-Yau might have some $\Pi_i = 0$, the overall volume is stabilized at a positive value. In such cases it could still happen that string-loop or other $\alpha'$-corrections shift the minimum to a point at which all four-cycles are positive and the overall volume is roughly the same. 

Consequently, we might now worry about the size of the flux density
\begin{equation}
\rho_{flux}=\frac{1}{\alpha'}\left(\int d^6y G_{3}\cdot \bar G_{3}
\right)^{1/2}\sim \frac{W_0}{\cal V}\quad.
\end{equation}
While supersymmetric flux stabilization of the type IIB axio-dilaton and the complex structure moduli has vanishing F-terms $D_\tau W = D_{a} W=0$ which removes their contribution to the flux density (see e.g.~section 2.3 in~\cite{Cicoli:2013swa}), volume stabilization requires the $(0,3)$-piece of $G_{3}$ to be non-zero in order to generate the VEV for $W$ in the first place. Hence, the F-terms of the K\"ahler moduli still produce a flux density
\begin{equation}
\rho_{flux}\sim \left(e^KK^{i\bar\jmath}D_{i}W_0\bar{D}_{\bar{j}}\bar{W}_0\right)^{1/2}\sim  \frac{W_0}{\cal V}
\end{equation}
scaling the same way as the naive ten-dimensional estimate above. However, inserting the scaling of the volume in our vacuum, we note that the fraction
$\lvert W_0 \lvert / \mathcal{V} \sim \lvert W_0 \lvert^{-2}$ and so one expects the flux density to actually decrease with increasing $W_0$ -- quite contrary to the situation known for KKLT or LVS class vacua.

In the minimum eq.~\eqref{taumin} the value of the cosmological constant is negative. To lift this vacuum to a metastable dS one may introduce an uplifting sector in the same way as it is done for LVS. We do not see any obstacles to an uplifting, since supersymmetry is already broken for eq.~\eqref{taumin}. 
\subsection{Estimating the Size of the $\Pi_i$ - a Simple Explicit Example}
\label{examplealexander}
At this point we have established the functional form of the contribution from the higher-derivative correction to the scalar potential. Moreover, we know that the positivity of the 2nd Chern class guarantees the positive semi-definiteness of its expansion coefficients $\Pi_i$ when using proper K\"ahler cone variables. In closing our discussion, we should like to have a ballpark estimate of the size of the $\Pi_i$ in order to assess the generic size of the new correction. 

To this end, we will provide results for the coefficients $\Pi_i$ in the expression $\int c_2\wedge J=\Pi_it^i$ for the well-known complete-intersection CY manifold $X_3=\mathbb{P}^4_{11169}\left[\begin{array}{c}18 \\ 4\end{array}\right]$ which has $h^{11}=2$, $h^{21}=272$, and consequently $\chi=-540<0$. This example was presented in \cite{Candelas:1994hw, Denef:2004dm, Balasubramanian:2005zx} and is of the ``Swiss-Cheese'' type. While this example cannot show volume stabilization due to its negative Euler number, its mirror does stabilize all the volumes, and we use the $\chi<0$ manifold just as an illustrative example to provide an estimate for the numerical size of the $\Pi_i$.

We can describe $X_3$ as the vanishing locus of the polynomial
\beq\label{CYeq}
\xi^2=P_{18,4}(u_i)
\eeq
in the ambient toric variety
\begin{equation}
X_4^{\text{amb}} :\quad
\begin{array}{cccccc}
u_1 & u_2 & u_3 & u_4 & u_5 & \xi\\ 
\hline 
1 & 1 & 1 & 6 & 0 & 9\\
0 & 0 & 0 & 1 & 1 & 2\\
\end{array}\quad .
\label{WSPex}
\end{equation}
Eq.~\eqref{CYeq} arises in Sen's limit as the double cover of the base $B_3=\mathbb{P}^1\to \mathbb{P}^2$ with twist $n=-6$ of an elliptically fibred CY 4-fold $Y_4:T_2\to B_3$.
From the above weight system data we can compute the linear relations and triple intersections of the toric divisors $D_{u_i}, D_{\xi}$ (given by the vanishing loci $u_i=0,\xi=0$), and their restriction to the hypersurface equation (see e.g.~\cite{Louis:2012nb}). This allows us to compute the total Chern class of $X_3$ by adjunction in terms of the Chern class of the embedding toric variety and the normal bundle of the hypersurface. Expanding to second order, we get the second Chern class of $X_3$ in terms of the elements of a basis of toric divisors. Carefully expanding the K\"ahler form $J=t^i D_i$ into a basis of divisors spanning the K\"ahler cone, i.e. where all $t_i\geq0$ simultaneously and independently from each other, we can then compute $\int c_2\wedge J$ using the known divisor triple intersection numbers on our CY 3-fold. Following the conventions of~\cite{Louis:2012nb}, we write $J=t^1D_1+t^5D^5$ and the Chern class computation produces
\beq
\int\limits_{X_3} c_2\wedge J=36t^1+102t^5\quad.
\eeq
Hence, this example served us to verify that the $\Pi_i \geq 0$, and provides us with a first estimate of their typical size to be ${\mathcal O}(10\ldots 100)$.
\newpage
\section{Conclusion}\label{sec:conclusions}
In the first part of this paper we revisited the ghost-free four-derivative sector for chiral superfields in $\mathcal{N}=1$ global supersymmetry as well as supergravity in superspace. This sector is captured by the operator in eq.~\eqref{eq:higher_derivative_global_sec2}. This term does not lead to a propagating auxiliary field, but induces cubic polynomial equations for the chiral auxiliaries and, thus, up to three inequivalent on-shell theories. We showed that within the context of effective field theory there is a unique physical on-shell theory, namely the theory with analyticity in the coupling $T$. The additional theories can be regarded as mere artefacts of a truncation of an infinite-series of higher-derivative operators in superspace, as was illustrated explicitly by the one-loop Wess-Zumino model in sec.~\ref{sec:appendix_wess_zumino}. This example furthermore revealed that the non-analytic theories are incapable of reproducing the non-local, untruncated 'UV'-theory. In addition we have demonstrated that in a regime of small kinetic terms all on-shell Lagrangians obtained from eq.~\eqref{eq:higher_derivative_global_sec2} are ghost-free. After clarification of these conceptual issues we displayed the general on-shell theory in eq.~\eqref{sugraL}.

In the second part of this paper we analysed the correction to the scalar potential, which is generated by the operator in eq.~\eqref{eq:higher_derivative_global_sec2}, and the properties of the vacua of the theory. Firstly, in situations in which the ordinary, two-derivative theory possesses a supersymmetric minimum, this minimum persists unchanged in the higher-derivative theory in agreement with the general discussion in \cite{Cecotti:1986jy}. If, one the other hand, supersymmetry was already broken in the two-derivative theory, then the higher-derivative operator might be of interest, specifically in situations, in which flat directions exist within the minimum. Unless a symmetry is protecting this flat direction or the flat direction is a Goldstone boson, as for example if supersymmetry breaking occurs via R-symmetry breaking, we expect that in general the higher-derivative operator lifts the flatness. For the case of global supersymmetry this was exemplified using the O'Raifeartaigh model. Within supergravity we provided a simple one-dimensional no-scale type model as a first example in sec.~\ref{sec:susy_flat_dir}.

Of special interest are theories, which do not have a minimum at two-derivative level. This is for instance the case for the K\"ahler moduli sector of type IIB flux compactifications on Calabi-Yau orientifolds after inclusion of the leading order $\alpha'$-corrections to the K\"ahler potential, but ignoring non-perturbative effects. 
We extended the analysis of how $(\alpha')^3$-corrections in ten dimensions modify the four-dimensional theory obtained after compactification to the higher-derivative sector. Specifically we found that corrections to the scalar potential, which are induced by terms with four powers of the flux three-form $G_3$ and which cannot be described via corrections to $K$ or $W$, fit into the off-shell operator in eq.~\eqref{eq:lagrangian_superspace_hd}. The respective four-derivative terms for the K\"ahler moduli can be found by reducing the ten-dimensional $R^4$ corrections. 
Contrary to the terms quartic in $G_3$, the $R^4$ term is fully known \cite{Antoniadis:1997eg, Antoniadis:2003sw} and, thus, we computed the four-derivative terms and inferred the correction to the scalar potential by matching to eq.~\eqref{sugraL}. The result is displayed in eq.~\eqref{V1twoB}. In this computation we omitted numerical factors. A proper treatment of these factors lies outside the scope of this paper as this requires a systematic understanding of the off-shell higher-derivatives in four dimensions. Notably (ghost-like) operators exist, which do not modify the scalar potential, but induce four-derivative terms of the same type as those obtained from the $R^4$ correction. However, it is important to note that eq.~\eqref{eq:lagrangian_superspace_hd} is the only off-shell operator which receives four-derivative terms from $R^4$ and contributes to the scalar potential at order $(\alpha')^3$, as we will demonstrate elsewhere. In other words, the respective correction to the scalar potential has a unique off-shell matching. Moreover, we neglected warping effects. In principle, warping-induced contributions are expected to enter the scalar potential at the same order in powers of inverse volume as the correction in eq.~\eqref{V1twoB}, for instance via terms $R^3 F_5^2$ in ten-dimensions. On the other hand it was recently shown that large cancellations associated with warping-induced terms occur in the context of $\alpha'$-corrections to the effective action of M-theory \cite{Grimm:2014efa}. Leading order warping-effects were also studied in \cite{Martucci:2014ska}. Thus, it will be interesting to test our approximation in the future. One might also worry whether the correction in eq.~\eqref{V1twoB} can be absorbed via a redefinition of the K\"ahler moduli, for instance in the spirit of \cite{Grimm:2013gma, Grimm:2013bha}. If this would be possible then also the contribution to the kinetic terms in eq.~\eqref{nonkahlercorrIIB} would be absorbed simultaneously. However, this term renders the metric multiplying the two-derivative term non-K\"ahler, as can be checked by computing the respective torsion, and, thus, such corrections are manifest.

In a second step we assessed whether the correction in eq.~\eqref{V1twoB} can lead to a theory with a minimum without taking into account non-perturbative effects. In sec.~\ref{MInimum} we indeed found that a model-independent minimum exists, where all four-cycle volumes are frozen to values which are determined by topological numbers encoded in the second Chern class, cf.~eq.~\eqref{taumin}. This holds for all Calabi-Yau threefolds with $\chi(M) > 0$ and provided that the undetermined overall numerical factor of the higher-derivative operator has a negative sign. This moduli stabilization scenario is intriguing as the structure and properties of the vacuum are determined purely from topological data of the Calabi-Yau and no additional ingredients are required. Moreover let us compare the vacuum in eq.~\eqref{taumin} and eq.~\eqref{Vmin} to the results of LVS. We obtain a minimum given that $\chi(M_3) > 0$, contrary to LVS, where it is necessary that $\chi(M_3) < 0$. To ensure a large volume in eq.~\eqref{Vmin} we see that a largish value $\lvert W_0 \lvert \,\gtrsim 1$ is preferred. The ensuing scaling of the stabilized value of the volume with the $W_0$ also renders both the 3-form flux density and the gravitino mass in eq.~\eqref{gravmass} small at large $W_0$. 

In the future it will be necessary to determine the sign of $\lambda$ to confirm the existence of the minimum in eq.~\eqref{taumin}. However, a prior systematic understanding of all higher-derivative operators in curved superspace is required. Furthermore, a better understanding of additional $(\alpha')^3$-corrections to the scalar potential, such as for instance warping-induced terms but also the subleading terms in inverse volume, is important in order to fully trust the minimum in eq.~\eqref{taumin}.

\section*{Acknowledgments}

We have benefited from conversations and correspondence with
Per Berglund, Shanta de Alwis, Emilian Dudas, Thomas Grimm, David Mattingly, Liam McAllister, Fabian R\"uhle and Lucila Zarate.

This work was supported by the German Science Foundation (DFG) under
the Collaborative Research Center (SFB) 676 Particles, Strings and the Early
Universe, and by the Impuls und Vernetzungsfond of the Helmholtz Association 
of German Research Centers under grant HZ-NG-603.

\newpage
\vskip 2cm
\appendix
\noindent
{\bf\Huge Appendix}
\section{Exact Solutions of the Cubic Equation for $F$}
\label{sec:appendix_explicit_solutions}
\subsection{One-Dimensional Models with Arbitrary $W$}
\label{sec:full_one_d_sol}
In this appendix we discuss the general solution of the equation of
motion for the auxiliary field $F$ of a single chiral multiplet in the context
of supergravity. All the results below can be extrapolated to the
case of global supersymmetry after reintroducing the factors of 
the Planck scale $M_p$ and performing the limit $M_p \rightarrow \infty$. 

Recall that the equation of motion for $F$ is cubic and given by
(cf.\ \eqref{eomauxiliary})
\begin{equation}\label{eomauxiliaryA1}
 F\left[\lvert F \lvert^2 +
    \mathrm{e}^{-K/3}\left({2T^{-1}}{G_{A\bar{A}}}-\lvert \partial A
       \lvert^2 \right)\right] + 2 T^{-1} \bar{D}_A \bar W =0\ .
\end{equation}
It is possible to rewrite \eqref{eomauxiliaryA1} as a holomorphic 
cubic equation with real coefficients after the 
field redefinition
\begin{equation}
 F = f(A,\bar A) \,\bar{D}_A \bar{W}\ ,
\end{equation}
where $f$ is the new auxiliary field variable 
and we assume $W\neq0$. Inserted
into \eqref{eomauxiliaryA1} we obtain
\begin{equation}\label{feq}
 f \left[ \lvert f \lvert^2 \,\lvert D_A W \lvert^2 
+ \mathrm{e}^{-K/3}\left(2 T^{-1}G_{A\bar{A}}-\lvert \partial A \lvert^2 \right) \right] + {2 T^{-1}} = 0 \;\;.
\end{equation}
Since $T$ and the expression in the square bracket
are real we see that also $f$  has to be real.
Therefore \eqref{feq} is of the form
\begin{equation}
\label{eq:cubic}
 f^3 +p f  + q = 0 \
\end{equation}
with 
\begin{equation}\begin{aligned}\label{pqdef}
 p = \frac{\mathrm{e}^{-K/3}}{\lvert D_A W
   \lvert^2}\left(\frac{G_{A\bar{A}}}{2 T}-\lvert \partial A
    \lvert^2 \right) \ ,\qquad
 q = \frac{1}{2 T\lvert D_A W \lvert^2}\;\;.
\end{aligned}\end{equation}
In the case of global supersymmetry it necessary to note that the
K\"{a}hler potential has mass dimension two, so that in the limit $M_p
\rightarrow \infty$ we get
\begin{equation}
  p \rightarrow \frac{1}{\lvert  W_{,A}
  \lvert^2}\left(\frac{G_{A\bar{A}}}{2 T}-\lvert \partial A
   \lvert^2 \right), \qquad
 q \rightarrow \frac{1}{2 T\lvert W_{,A} \lvert^2}\;\;.
\end{equation}

Eq.~\eqref{eq:cubic} 
is a cubic equation with real coefficients $p,q$ and its solutions are
known. However, in general only one out of the three possible
solutions is real. There are different regimes of interest \cite{Gwyn:2014wna}: 

\begin{enumerate}[(1)]
 \item $p>0$: In this case only one real solution exists given by
\begin{equation}
 \label{eq:sol_reg1}
 f_{(1)} = -2\sqrt{\frac{p}{3}} \,\mathrm{sinh}\left[\frac{1}{3}\mathrm{arsinh}\left(\sqrt{x}\right) \right] \;\;,
\end{equation}
where we defined
\begin{equation}\label{xdef}
 x \equiv \frac{27 q^2}{4p^3}  \;\;.
\end{equation}
\item $p<0$ and $4p^3+27q^2>0$: Here also only one real solution exists, which reads
\begin{equation}
\label{eq:sol_reg2}
 f_{(2)} = -2\,\mathrm{sign}(q)\,\sqrt{-\frac{p}{3}} \,\mathrm{cosh}\left[\frac{1}{3}\mathrm{arcosh}\left(\sqrt{-x}\right) \right] \;\;,
\end{equation}
\item $p<0$ and $4p^3+27q^2<0$: In this regime all three solutions are real and can be expressed as follows
\begin{equation}
\label{eq:sol_reg3}
 f_{(3),k} = 2\,\sqrt{-\frac{p}{3}} \,\mathrm{cos}\left[\frac{1}{3}\mathrm{arccos}\left(\sqrt{-x}\right)-\frac{2\pi \,k}{3} \right] \;\;\;\;, \;\;\;\;k = 0,1,2 \;\;.
\end{equation}
\end{enumerate}
In terms of the variable $x$ defined in \eqref{xdef}
the different regimes can be expressed by
\begin{equation}
\label{eq:dif_regimes}
 (1): \; x>0 \ , \qquad (2) : \; x<-1 \ ,\qquad (3) : \; -1 < x < 0\ .
\end{equation}

Let us make a few remarks regarding the different regimes. Suppose
that $T$ is a constant. Then, for $T<0$ we have $p<0$ and so one is
always in regime (2) or (3). For simplicity let us assume that the
contribution of the kinetic terms in $p$ is negligible, then the
difference between the regions is characterized by 
\begin{equation}
 27 \lvert D_A W \lvert^2 \,\gtrless -2T^{-1}\mathrm{e}^{-K}
   G_{A\bar{A}}^3\ .
\end{equation}
If we take the kinetic contribution into account, one can directly see
that for large kinetic terms $p$ becomes large and negative so that
one always reaches regime (3). For $T>0$ one can be in all three
regions.
Note, that if we would restrict ourselves to the discussion of the non-derivative component of $F$, the condition $T>0$ could only be supported in region (1). Moreover, we see that the different regions are dynamically connected. For instance a theory with $T>0$ could describe a dynamical field with initially small kinetic terms, thus, being described by the appropriate theory in regime (1). However, it could be that the kinetic terms are growing with the evolution of the field and hence one reaches regimes (2) and finally (3).
\subsection{Analysis of Kinetic Terms}
\label{sec:appendix_ghosts}
In this appendix we will demonstrate the absence of ghosts in the on-shell theories in all three regimes. 
To this end we compute the sign of the ordinary kinetic term in the on-shell Lagrangian. We will conduct the 
analysis in the context of supergravity and the results extrapolate directly to the case of global supersymmetry. 
Eliminating the auxiliary field $F$ from \eqref{sugrawf}
and keeping only terms which contribute to the standard kinetic term we arrive at
\begin{equation}
 \frac{\mathcal{L}}{\sqrt{-g}} \supset -G_{A\bar{A}} \lvert \partial A\lvert^2  + f\, \mathrm{e}^{2K/3}\, \lvert D_A W \lvert^2 - T \,\mathrm{e}^{2K/3} \,f^4\, \lvert D_A W \lvert^4 \;\;,
\end{equation}
where $f$ was determined in the previous section and via \eqref{pqdef}
depends on $\partial A$.
Let us expand the above terms in $\lvert \partial A \lvert^2$ assuming
that they are sufficiently small. 
The coefficient $\Sigma$ of the first term in the expansion determines the sign of the ordinary kinetic term. Making use of \eqref{eq:cubic} $\Sigma$ is given by
\begin{equation}
\mathcal{L} =
 \Sigma G_{A\bar{A}} \lvert \partial A\lvert^2\sqrt{-g} +\ldots\ ,\qquad
\Sigma = 
 -\left[1 + \frac{\partial f}{\partial p}\Bigl|_{0}\left(2f_0 + 3\frac{p_0}{q}\right)\right] \;\;.
\end{equation}
Here the subscript zero denotes that a quantity  is evaluated at
$\lvert \partial A \lvert^2=0$. Using the solutions for $f$ in the
three regimes given in eqs.~\eqref{eq:sol_reg1} to \eqref{eq:sol_reg3}
it is always possible to express $\Sigma$ as a function of $x_0$
(defined in \eqref{xdef}) only. More precisely, in each regime one finds the following:
\begin{enumerate}[(1)]
 \item $x>0$: One obtains
\begin{align}
 \nonumber \Sigma_{(1)} = -\Biggl\{ 1 + &\left[-\mathrm{sinh}\left(\frac{1}{3}\mathrm{arsinh}(\sqrt{x_0})\right)+\sqrt{\frac{x_0}{1+x_0}}\mathrm{cosh}\left(\frac{1}{3}\mathrm{arsinh}(\sqrt{x_0})\right)\right] \\
 & \times \left[-\frac{4}{3}\mathrm{sinh}\left(\frac{1}{3}\mathrm{arsinh}(\sqrt{x_0})\right)+\frac{2}{3}\sqrt{x_0}\right] \Biggr\}
\end{align}
Inspecting \eqref{pqdef} and \eqref{xdef} one finds that in this
regime 
$T>0$ has to hold in order to ensure $x>0$. Thus, we necessarily also
have $x_0 >0$ and then numerical evaluation shows
that $\Sigma$ is always negative  implying that this region is ghost-free. 
\item $x<-1$: The computation in this case has to be done more
   carefully, since $x<-1$ can occur for $T<0$ and $T>0$. From
   \eqref{pqdef}
we see that in the latter case the kinetic term has to be large and an
expansion around zero is not meaningful. For $T<0$ we have that $x_0 <0$
and computing $\Sigma$ yields
\begin{align}
 \nonumber \Sigma_{(2)} = -\Biggl\{ 1 + &\left[\mathrm{cosh}\left(\frac{1}{3}\mathrm{arcosh}(\sqrt{-x_0})\right)+\sqrt{\frac{x_0}{1+x_0}}\mathrm{sinh}\left(\frac{1}{3}\mathrm{arcosh}(\sqrt{-x_0})\right)\right] \\
 & \times \left[\frac{4}{3}\mathrm{cosh}\left(\frac{1}{3}\mathrm{arcosh}(\sqrt{-x_0})\right)+\frac{2}{3}\sqrt{-x_0}\right] \Biggr\} \;\;.
\end{align}
 $ \Sigma_{(2)}$  is discontinuous at $x_0 = -1$, but numerical
 evaluation shows
that it is negative for all $x_0 < 0$, which again implies the absence
of ghosts. 
\item $-1<x<0$: Here one finds that
\begin{align}
  \nonumber \Sigma_{(3),k} = -\Biggl\{ 1 + &\left[\mathrm{cos}\left(\frac{1}{3}\mathrm{arccos}(\sqrt{-x_0})-\frac{2\pi \,k}{3} \right)+\sqrt{\frac{-x_0}{1+x_0}}\mathrm{sin}\left(\frac{1}{3}\mathrm{arccos}(\sqrt{-x_0})-\frac{2\pi \,k}{3} \right)\right] \\
 & \times \left[\frac{4}{3}\mathrm{cos}\left(\frac{1}{3}\mathrm{arccos}(\sqrt{-x_0})-\frac{2\pi \,k}{3} \right)-\mathrm{sign}(T)\,\frac{2}{3}\sqrt{-x_0}\right] \Biggr\} \;\;.
\end{align}
Again there are two cases to discuss: For $T<0$, we always have $-1<x<x_0<0$ and all branches are ghost-free. On the other hand for $T>0$ we have that $x_0>0$. It is not expected that $\Sigma$ is defined here, since this corresponds to large $\lvert \partial A \lvert^2$.
\end{enumerate}

Let us make some additional comments about the appearance of ghosts in
those theories, where we truncate the theory to linear order in
$T$. 
In region (1), the auxiliary field is analytic in $T$ and
the lowest order contributions to the Lagrangian generated by the
auxiliary can be obtained as in \eqref{eq:analytic_branch}
and are given by\footnote{In contrast to the previous appendices here we introduced additional factors of the K\"{a}hler metric according to eq.~\eqref{eq:coupling_tensor_metric}.}
\begin{equation}
\label{eq:analytic_branch_for_L}
 \mathcal{L}_{(1)} \supset - \mathrm{e}^K G^{A \bar{A}} \lvert
 D_A W \lvert^2 + T\left[ \left(\mathrm{e}^K G^{A \bar{A}} \lvert
       D_A W \lvert^2 \right)^2  - 2 G_{A \bar{A}}
    \lvert \partial A \lvert^2  \mathrm{e}^K G^{A \bar{A}} \lvert
    D_A W \lvert^2 \right] + \mathcal{O}(T^2)\ .
\end{equation}
In region (2) on the other hand the auxiliary field has a pole at $T \rightarrow 0$ and hence is not analytic. The respective contributions to the Lagrangian are of the form
\begin{equation}
 \mathcal{L}_{(2)} \supset - 4T^{-1}+ \left( \frac{1}{2}
    \mathrm{e}^K G^{A \bar{A}} \lvert D_A W \lvert^2 +
    G_{A\bar{A}} \lvert \partial A \lvert^2 \right) +
 \mathcal{O}(\sqrt{T})\ .
\end{equation}
The fact that there exists a region in which the Lagrangian is not
analytic in $T$ is not surprising, since the limit $T \rightarrow 0$
with fixed $K,W$ automatically implies that one must exit region $(2)$
and enter region $(3)$
as can be seen from \eqref{pqdef}. In the third region the Lagrangian coincides with $\mathcal{L}_{(2)}$ for $k=0,2$ and with $\mathcal{L}_{(1)}$ for $k=1$.

Inspecting $\mathcal{L}_{(1)}$ we observe that the theory becomes ghost-like, once 
\begin{equation}
 2 \left( \mathrm{e}^K G^{A \bar{A}} \lvert D_A W \lvert^2 \right)^2 T \lesssim -1 \;\;.
\end{equation}
Equivalently this reads $x_0 \lesssim - 27/4$. However, such values of
$x_0$ correspond to regime $(2)$, where no analytic Lagrangian
exists. This indicates that the expansion of the analytic solution
fails to converge around values where the theory becomes ghostlike. In
the above we have treated the solutions to all orders in $T$. This is
sensible only, if we know that all higher-order contributions to the
EAFP vanish. However, even in the situation where we know the EAFP
only up to four-derivative level, we expect that the theory
becomes unreliable near $x_0 \sim -1$, which coincides with the
threshold, where the kinetic term starts to behave ghostlike.

\subsection{Analytic Solution in Arbitrary Dimensions}
\label{app:fullsol}
To complete the discussion of the four-derivative operator let us
analyse eq.~\eqref{eomauxiliary} for arbitrary
dimension. If we assume that the coupling tensor is given by
eq.~\eqref{eq:coupling_tensor_metric} and we only look for analytic
solutions, then the task is feasible and the answer unique. Inspecting
the equations of motion for the $F^i$  shows  that all analytic solutions have to be of the form
\begin{equation}
 F^i = \mathrm{e}^{K/3} G^{i\bar{j}} \bar{D}_{\bar j}\bar W f \;\;,
\end{equation}
where $f$ is analytic in $T$.\footnote{One can directly show via
  induction that the solution must reduce to this by assuming a
  general analytic expansion in $T$.} 
Inserted into eq.~\eqref{eomauxiliary} yields
\begin{equation}
 2 \lvert f \lvert^2 f T (\mathrm{e}^K G^{i\bar{j}} D_i W
 \bar{D}_{\bar j} \bar W) + f + 1 =0\ ,
\end{equation}
where for simplicity we ignore the dependence on the derivatives of
the scalar fields. As in section \ref{sec:full_one_d_sol}
we obtain the additional condition that $f$
has to be real-valued and, hence, the cubic equation reduces to
\begin{equation}
 f^3 + pf +p =0 \ , \qquad \mathrm{where}\qquad p =
(2T\mathrm{e}^K G^{i\bar{j}} D_i W \bar{D}_{\bar j}\bar W)^{-1}\ .
\end{equation}
Thus, the exact solution is be given by
\begin{equation}
\label{eq:F_mod_stab_full_sol}
 F^i = \mathrm{e}^{K/3} G^{i\bar{j}} \bar{D}_{\bar j}\bar W
 \sqrt{-\frac{4p}{3}}\,\mathrm{cos}\left[\frac{1}{3}\mathrm{arccos}\left(\sqrt{-\frac{27}{4p}}
    \right)-\frac{2\pi}{3} \right]\ .
\end{equation}
\section{Higher-Derivatives for K\"ahler Moduli from String-Theoretic $\alpha'$-Corrections}
\label{app:tensoralphaprime}
In this appendix we discuss how to compute four-derivative terms for K\"ahler class deformations from $(\alpha')^3 R^4$ corrections to the action of IIB in the context of flux compactifications on Calabi-Yau orientifolds. These corrections were already presented in eq.~\eqref{J0}. Notably $J_0$ generates the $\hat{\xi}$-correction to the K\"ahler potential in eq.~\eqref{eq:kahler_moduli_sugra} as shown in \cite{Becker:2002nn}. The following derivation is many ways analogous to the computation in this reference.

Before turning to the explicit analysis let us stress that we will focus on obtaining the overall functional form of the coupling tensor and omit the details of numerical factors. A proper treatment of these factors lies outside the scope of this paper as a complete discussion of the four-derivative bosonic action is required. 
Specifically a full understanding of all off-shell operators in four dimensions is necessary, that contribute four-derivative terms for the scalar fields. For instance off-shell higher-derivative operators exist, which mix with $(\partial \mathcal{V})^4$, but do not correct the scalar potential. Note that the operator in eq.~\eqref{eq:lagrangian_superspace_hd} is the only higher-derivative operator, that receives four-derivative terms from $J_0$ and can contribute to the scalar potential at order $\mathcal{O}(\alpha'^3)$, as we will demonstrate elsewhere.

The necessary terms of the 10-dimensional string-frame action for this appendix are given by\footnote{For brevity we do not display the terms for the RR and NSNS field strength forms here.} \cite{Antoniadis:1997eg}
\begin{equation}\label{S10}
 S_{(10)} \supset -\frac{1}{\kappa_{10}^2} \int \mathrm{d}^{10} x \sqrt{-g^{(10)}} \mathrm{e}^{-2\phi} \left(R + 4(\partial \phi)^2 + \frac{(\alpha')^3 \zeta(3)}{3 \cdot 2^{11}} J_0 \right) \ .
\end{equation}
Here $R$ denotes Ricci scalar, $\phi$ the ten-dimensional dilaton, $g^{(10)}$ the metric and $J_0$ was given in eq.~\eqref{J0}.\footnote{Note that the dilaton receives higher-derivative corrections \cite{Kehagias:1997cq}. In the following, we shall consider the dilaton only at the two-derivative level and hence stick to the results of \cite{Becker:2002nn}.} There exists a basis of 26 independent contractions of four Riemann tensors, in which $J_0$ necessarily has to expand \cite{Fulling:1992vm}. Here we do not compute the exact coefficients of this expansion, but simply argue within this basis of the 26 terms to obtain the functional form of the possible four-derivative terms. 

We will not compute the coupling of gravity to the higher-derivatives of the K\"ahler moduli and, therefore, set the four-dimensional piece of the metric to a Minkowski-form. Furthermore we will neglect the warping-factor, which is non-trivial in the presence of background-fluxes. For simplicity we will conduct the analysis with a single K\"ahler-type deformation turned on. Altogether the ten-dimensional metric then reads
\begin{equation}\label{10dmetric}
 \d s_{(10)}^2 = g_{MN} \d x^M \d x^N = \eta_{\mu \nu} \d x^{\mu} \d x^{\nu} + g_{mn} \d y^m \d y^n \ , 
\end{equation}
where $M,N=0,\ldots,9$, the $y^m, m=1,\ldots,6$ are real coordinates on the compact manifold $M_3$ and 
$g_{mn} = \mathrm{e}^{2u(x)} \widetilde{g}_{mn}(y)$.
The volume measured by the background metric $\widetilde{g}_{mn}$ is
normalized to unity, i.e.\ we choose $(2\pi\alpha')=1$. 
This way the Planck constants in ten and four
dimensions can be directly related to each other as $\kappa_{10}^{-2}
= \kappa_4^{-2}$. The single volume modulus in the string-frame is 
normalized as $\mathrm{e}^{6u} = \hat{\mathcal{V}}$. 

Note that the higher-curvature terms in eq.~\eqref{S10} 
modify the Einstein equations. More precisely the 
Einstein equations along the internal directions read \cite{Freeman:1986}
\begin{equation}\label{modEE}
 R_{\alpha \bar{\beta}} \sim (\alpha')^3 \partial_{\alpha} \partial_{\bar{\beta}} Q \ ,
\end{equation}
where we introduced local complex coordinates 
$(z^{\alpha}, \bar{z}^{\bar{\beta}})$ with $\alpha, \bar\beta = 1,2,3$ 
on the internal manifold. Furthermore $Q$ denotes the six-dimensional 
Euler integrand, i.e.~$\int d^6 y \sqrt{g} Q = \chi(M_3)$. As a
consequence of eq.~\eqref{modEE} the background metric in eq.~\eqref{10dmetric} 
is in general not Ricci-flat. Eq.~\eqref{modEE} is formally solved by 
\begin{equation}
 \widetilde{g}_{mn} = \widetilde{g}_{mn}^{\,(0)} + (\alpha')^3 \widetilde{g}_{mn}^{\,(1)} \ ,
\end{equation}
where $\widetilde{g}_{mn}^{\,(0)}$ is a Ricci-flat metric solving the zeroth-order 
Einstein equations and $\widetilde{g}_{mn}^{\,(1)}$ solves eq.~\eqref{modEE} 
at order $(\alpha')^3$. When reducing $J_0$ it is not necessary to take into account the 
correction $\widetilde{g}_{mn}^{\,(1)}$ as it enters at order $(\alpha')^6$. 
The leading $(\alpha')^0$ terms in eq.~\eqref{S10} on the other hand 
induce $(\alpha')^3$-corrections via $\widetilde{g}_{mn}^{\,(1)}$. However, 
the respective correction coming from the standard Einstein-Hilbert term 
is a total derivative. The remaining terms do not correct the kinetic terms.
Hence, we will in the following ignore the correction $\widetilde{g}_{mn}^{\,(1)}$ 
and treat $\widetilde{g}_{mn}$ as Ricci-flat.

To determine the curvature terms inside $J_0$ we need to compute the components of the Riemann tensor. In the following we use the conventions 
\begin{equation}\begin{aligned}
 R^M {}_{NPQ} &= \partial_P \Gamma^M_{QN} - \partial_Q \Gamma^M_{PN} + \Gamma^R_{QN} \Gamma^M_{PR} - \Gamma^R_{PN} \Gamma^M_{QR} \ ,\\
 \Gamma^M_{PN} &= \frac{1}{2} g^{MQ} \left(\partial_{P} g_{NQ} + \partial_{N} g_{PQ} - \partial_Q g_{PN} \right) \ .
\end{aligned}\end{equation}
Up to symmetries there are only two non-vanishing independent pieces of the Riemann tensor computed with respect to the metric in eq.~\eqref{10dmetric}. They are given by
\begin{equation} \begin{aligned}
 R_{m \mu n \nu} &= -g_{mn} (\partial_{\mu} u \partial_{\nu} u + \partial_{\mu} \partial_{\nu} u ) \ ,\\
 R_{k m n p} &= \mathrm{e}^{2u} \widetilde{R}_{kmnp} + (\partial u )^2 (g_{kp}g_{mn}-g_{kn}g_{pm}) \ .
\end{aligned}\end{equation}
Here $\widetilde{R}_{kmnp}$ denotes the Riemann tensor components of the background metric $\widetilde{g}_{mn}$. From the Riemann tensor we can furthermore compute the Ricci-tensor as well as the scalar curvature
\begin{equation} \begin{aligned}
 R_{\mu \nu} &= -6(\partial_{\mu} u \partial_{\nu} u + \partial_{\mu} \partial_{\nu} u ) \ ,\quad
 R_{m n} = - g_{mn} ( 6(\partial u )^2 + \Box u) \ ,\quad
 R = -42 (\partial u )^2 - 12 \Box u\ .
\end{aligned}\end{equation}
It is evident that in the reduction of eq.~\eqref{J0} one obtains terms with up to eight derivatives of $u$. Here we are solely interested in the terms with four-derivatives. Computation of all 26 basis elements in \cite{Fulling:1992vm} shows that one obtains the following four-derivative terms
\begin{equation}\begin{aligned} \label{J02}
 J_0 \supset  \mathrm{e}^{-4 u} \Bigl[ &\alpha_1 (\partial u)^4 + \alpha_2 \Box u (\partial u)^2 + \alpha_3 (\Box u)^2 + \alpha_4 (\partial_{\mu} \partial_{\nu} u)(\partial^{\mu} \partial^{\nu} u) \\
 &+ \alpha_5 (\partial_{\mu} \partial_{\nu} u)(\partial^{\mu}u)(\partial^{\nu} u) \Bigr]\widetilde{R}_{kmnp} \widetilde{R}^{kmnp}\ ,
\end{aligned}\end{equation}
for some constants $\alpha_i$. Since for a Calabi-Yau $\widetilde{R}_{mn} = 0$, the only non-vanishing contraction of two Riemann tensors is given by $\widetilde{R}_{kmnp} \widetilde{R}^{kmnp}$. We see that five different four-derivative terms appear here. However, in the four-dimensional action these terms are not independent and related by partial integration.\footnote{For example partial integration reveals that $(\partial_{\mu} \partial_{\nu} u)(\partial^{\mu}u)(\partial^{\nu} u)$ can be recast into a combination of $ (\partial u)^4$  and $\Box u (\partial u)^2 $.} In the following we confine the discussion to the first term in eq.~\eqref{J02}. This is justified as we merely wish to obtain functional dependencies rather than explicit coefficients. The other four-derivative terms induce ghost-like degrees of freedom and are not of interest here. It is convenient to express the Riemann-tensor square with respect to $g_{mn}$ again. Up to derivatives we have
\begin{equation}
 {R}_{kmnp} {R}^{kmnp} =  \mathrm{e}^{-4 u} \widetilde{R}_{kmnp} \widetilde{R}^{kmnp} + \dots
\end{equation}
In the action we obtain at order $(\alpha')^3$
\begin{equation}
\label{pieceofaction}
 S_{(\partial u)^4} = - \frac{1}{2\kappa_4^2} \int \d^4 x \sqrt{-g} \,\mathrm{e}^{-2\phi_0} \,\alpha_1\, (\partial u)^4 \int_{M_3} \d^6 y \sqrt{g} \,{R}_{kmnp} {R}^{kmnp} \ .
\end{equation}
It is convenient to rewrite the integral over the compact dimensions as follows
\begin{equation}
\label{c2wedgeJ}
 \int_{M_3} \d^6 y \sqrt{g} \,{R}_{kmnp} {R}^{kmnp} \sim \int_{M_3} c_2 \wedge J \ ,
\end{equation}
where $c_2$ is the second Chern class of the Calabi-Yau threefold and $J$ its K\"ahler form. This can be checked directly using local complex coordinates. With respect to these coordinates we have
\begin{equation}\begin{aligned}
 c_2 &= \frac{1}{2} \left( \mathrm{Tr}\mathcal{R}^2 - (\mathrm{Tr}\mathcal{R})^2 \right) \ ,\qquad
  J &= i g_{\alpha \bar{\beta}} \d z^{\alpha} \wedge \d \bar{z}^{\bar{\beta}} \ ,
\end{aligned}\end{equation}
where $\mathcal{R}$ is the curvature two-form.
The traces of the curvature two-form are given by
\begin{equation}\begin{aligned}
 \mathrm{Tr}\mathcal{R} &= R^{\alpha} {}_{\alpha \beta \bar{\gamma}} \, \d z^{\beta} \wedge \d \bar{z}^{\bar{\gamma}} \ ,\qquad
 \mathrm{Tr}\mathcal{R}^2 &= R^{\alpha} {}_{\beta \gamma \bar{\delta}} R^{\beta} {}_{\alpha \epsilon \bar{\zeta}} \, \d z^{\gamma} \wedge \d \bar{z}^{\bar{\delta}} \wedge  \d z^{\epsilon} \wedge \d \bar{z}^{\bar{\zeta}}\ .
\end{aligned}\end{equation}
On a Calabi-Yau the first Chern class vanishes and, hence, we have $\mathrm{Tr}\mathcal{R}=0$.\footnote{To prove eq.~\eqref{c2wedgeJ} it is also helpful to note the relation $\sqrt{\mathrm{det}(\widetilde{g}_{mn})} = \mathrm{det}(g_{\alpha \bar{\beta}})$, which links the volume forms of the two different coordinate charts to each other.}

From eq.~\eqref{c2wedgeJ} it is evident that
\begin{equation}
 \int_{M_3} c_2 \wedge J \geq 0 \ .
\end{equation}
Here equality holds, if and only if $M_3$ has constant holomorphic sectional curvature \cite{CHEN01011975}. For K\"{a}hler manifolds with constant holomorphic sectional curvature $c$ the Riemann tensor must necessarily take the form \cite{kobayashi1969foundations}
\begin{equation}
 R_{\alpha \bar{\beta} \gamma \bar{\delta}} = - \frac{c}{2} \left( g_{\alpha \bar{\beta}} g_{\gamma \bar{\delta}} + g_{\alpha \bar{\delta}} g_{\gamma \bar{\beta}}\right)
\end{equation}
and, thus, for Calabi-Yau manifolds $c=0 = R_{\alpha \bar{\beta} \gamma \bar{\delta}}$. This is only possible if $M_3$ is a torus $T^6$.

The term in eq.~\eqref{pieceofaction} is expressed in the string frame. In order to transform to the Einstein frame note that the two-derivative part of the bosonic action is given by \cite{Becker:2002nn}\footnote{We promote $\eta_{\mu\nu}$ to an arbitrary Lorentzian metric $g_{\mu\nu}$ here.}
\begin{equation}\begin{aligned}
 S = - \frac{1}{2\kappa_4^2} \int &\d^4 x \sqrt{-g} \,\mathrm{e}^{-2\phi_0} \left( \mathrm{e}^{6u} + \frac{\xi}{2} \right) R^{(4)} +\ldots
\end{aligned}\end{equation}
where 
$\xi$ parametrizes the leading $\alpha'$-corrections and is given in eq.~\eqref{xi} and $R^{(4)}$ denotes the scalar curvature in four dimensions.


The next step is to transform into the four-dimensional Einstein frame
via a Weyl rescaling. Simultaneously one has to rediagonalize the 
kinetic terms for the scalar fields. This is achieved by the
redefinitions
\begin{equation}\begin{aligned}\label{Weyltrans}
  g^{(E)}_{\mu\nu} &= \mathrm{e}^{-\phi_0/2}\left(\mathcal{V} + \frac{\hat\xi}{2} \right)
  g_{\mu\nu} \ ,\qquad
\mathcal{V}= \hat{\mathcal{V}}\mathrm{e}^{-3\phi_0/2} =
\mathrm{e}^{-3\phi_0/2}\mathrm{e}^{6u}\ .
\end{aligned}\end{equation}
where $\hat\xi$ is defined in \eqref{xi}. Here one observes that also couplings of 
the four-dimensional Riemann tensor to the K\"ahler deformation contribute 
to the four-derivative term for $u$ after the Weyl rescaling.\footnote{A coupling of the $\mathcal{N}=2$ vector multiplets to four-dimensional curvature invariants is forbidden by supersymmetry \cite{Antoniadis:1997eg} and, hence, one might expect these couplings also to be absent in the $\mathcal{N}=1$ sector. However, a coupling of the four dimensional Riemann-tensor to derivatives of the K\"ahler moduli might be present.}

Even though we considered only a single volume modulus so
far, in the following the results can be generalized to the
situation of arbitrarily many K\"ahler moduli. 
The proper $\mathcal{N}=1$
field variables are \cite{Becker:2002nn}
\begin{equation}\begin{aligned}\label{N=1coords}
 T^i &= \frac{1}{3}\left(g^i + i \mathcal{V}^i \right) \ ,\qquad
 \tau &= l + i \mathrm{e}^{-\phi_0} \ ,
\end{aligned}\end{equation}
where $l$ is the R-R scalar and $g^i$ originate from the R-R four-form. The imaginary components of $T^i$ are given by rescaled four-cycle volumes as follows
\begin{equation}\label{weylresc}
 \mathcal{V}^i = \tau_i = \frac{\partial \mathcal{V}}{\partial t^i} \
 ,\qquad \textrm{where} \qquad t^i = \hat{t}^i \mathrm{e}^{-\phi_0/2}
 \ , 
\end{equation}
denote the Einstein-frame two-cycle volumes. Furthermore $\hat{t}^i$ are the two-cycle volumes measured in the string-frame.
These are related to the overall volume via eq.~\eqref{eq:volume_of_cy}.
For a generic Calabi-Yau threefold we can expand $J = \sum_{i=1}^{h^{1,1}} \hat{t}^i \hat{D}_i$, where $\hat{D}_i$ form a basis of the Dolbeault cohomology $H^{1,1}(M_3,\mathbb{Z})$. Hence, the integral on the r.h.s. of eq.~\eqref{c2wedgeJ} can be understood as
\begin{equation}\label{c2wedJ}
 \int_{M_3} c_2 \wedge J = \hat{t}^i  \int_{M_3} c_2 \wedge \hat{D}_i  \equiv \Pi_i \, \hat{t}^i \ ,
\end{equation}
where $\Pi_i$ is a number encoding the topological information of the second chern class.

Up to terms involving derivatives of the dilaton we can use the above coordinates to rewrite eq.~\eqref{pieceofaction} 
\begin{equation}\begin{aligned} \label{pieceofaction4}
  S_{(\partial u)^4} \sim  - \frac{1}{2\kappa_4^2} \int & \d^4 x \sqrt{-g^{(E)}} \,(\Pi_m \, t^m)\left[ \frac{1}{2i} (\tau - \bar{\tau}) \right]^{3/2} \\
                             & \times  K_{(0),i} K_{(0),j} K_{(0),k} K_{(0),l} \,(\partial_{\mu} \tau_i  \partial^{\mu} \tau_j)( \partial_{\nu} \tau_k \partial^{\nu} \tau_l) \ ,
\end{aligned}\end{equation}
where $K_{(0)} = -2 \,\mathrm{ln}(\hat{\mathcal{V}})$ denotes the classical K\"ahler potential of the underlying $\mathcal{N}=1$ geometry. Finally we can match this result to the Lagrangian in eq.~\eqref{sugraL} and read off the coupling tensor
\begin{equation}
\label{resultTapp}
 T_{ij \bar{k} \bar{l}} = \lambda (\alpha')^3 (\Pi_m \, t^m) \left[\frac{1}{2i}(\tau - \bar{\tau})\right]^{3/2} K_{(0),i} K_{(0),j} K_{(0),\bar{k}} K_{(0),\bar{l}} \ ,
\end{equation}
where $\lambda$ denotes the overall unknown numerical factor. Its computation is beyond the scope of this paper as we discussed at the beginning of this appendix.

In the last steps we generalized to the case of arbitrarily many K\"ahler moduli even though we took into account only a single modulus during the compactification. 
When arbitrarily many K\"ahler-class deformations are switched on the coupling tensor might differ from eq.~\eqref{resultTapp}. For instance, just as for the ordinary kinetic term, the K\"ahler metric $K_{(0),i\bar{j}}$ could appear. Even though the coupling tensor computed for arbitrary $h^{1,1}$ could be different from eq.~\eqref{resultTapp}, there is evidence that the induced correction to the scalar potential 
can be inferred from the computation with $h^{1,1}=1$ without loss of generality. To see this we will make use of the results of appendix~\ref{app:geometrytensor}, which we will briefly summarize now. In the large volume limit the correction to the scalar potential in eq.~\eqref{V1gen} behaves as
\begin{equation}
\label{HDcorrV}
 V_{(1)} = - \frac{\lvert W_0 \lvert^4}{\mathcal{V}^4}\, {T_{(0)}}^{\bar{i}\bar{j} kl} K_{(0),\bar{i}} K_{(0),\bar{j}} K_{(0),k} K_{(0),l} + \dots\ ,
\end{equation}
where $T_{(0)}$ is the coupling tensor truncated to the leading order
term in the large volume limit. From the above index structure it is clear 
that $T_{(0)}$ is a tensor in the geometry defined by the K\"ahler potential 
$K_{(0)}$. We assume that its tensor structure is derived from $K_{(0)}$, which means
that any indexed quantity appearing within $T_{(0)}$ is related to derivatives of $K_{(0)}$ and possibly
contractions with the inverse K\"ahler metric, see appendix~\ref{app:geometrytensor} for more details. 
In appendix~\ref{app:geometrytensor} we study eq.~\eqref{HDcorrV} in detail 
and provide evidence for the following statement: If $T_{(0)}$ does not involve any scalar function 
and, hence, only consists of objects with at least one index, then $V_{(1)} \sim \mathcal{V}^{-4}$ up to
some constant. Thus, an additional dependence of $V_{(1)}$ upon $\mathcal{V}$ or $\tau_i$ 
can only be generated by scalar functions appearing within $T_{(0)}$.

When reducing $J_0$ with an arbitrary number of K\"ahler-type deformations turned on, the four-derivative terms are 
again obtained from those contractions where two out of the four Riemann tensors have indices along the internal 
directions and, thus, contribute a factor $\int c_2 \wedge J$. The remaining indices yield contracted metrics 
or derivatives. We infer that the general coupling tensor should be of the form
\begin{equation}
 T_{ij \bar{k} \bar{l}} \sim (\Pi_m \, t^m) \mathcal{T}_{ij \bar{k} \bar{l}}
\end{equation}
where $\mathcal{T}$ is a tensor, that consists purely of indexed quantities. 
As we consider only terms at order $(\alpha')^3$ this tensor is a tensor in the geometry 
defined by $K_{(0)}$. Thus, we can apply the results of the appendix~\ref{app:geometrytensor} and conclude 
that the functional behaviour of eq.~\eqref{HDcorrV} is captured by $\int c_2 \wedge J$, which
was already present in the computation with $h^{1,1}=1$.

\newpage

\section{K\"{a}hler Moduli Space and Coupling Tensor}
\label{app:geometrytensor}
In this appendix we study the correction to the scalar potential induced by the higher-derivative operator in eq.~\eqref{HDcorrV} for the geometry of the K\"{a}hler moduli at leading order in the large volume limit, that is for $K_{(0)} = -2\,\mathrm{ln}(\mathcal{V}), \, W=W_0$ and $\mathcal{V}$ given by eq.~\eqref{eq:volume_of_cy}.\footnote{Some of the below results can also be found in \cite{Covi:2008ea}.} Within this appendix we set $K_{(0)} = K$ and $T_{(0)} = T$ for brevity. Up to factors the relevant object of study is given by
\begin{equation}
 \mathcal{Z} \equiv T_{ij\bar{k}\bar{l}} K^i K^j K^{\bar{k}} K^{\bar{l}} \ ,
\end{equation}
where $K^i = K^{i\bar{j}} K_{\bar{j}}$ and $K^{i\bar{j}}$ denotes the inverse K\"{a}hler metric. 
Due to the shift-symmetry of $K$ in the following we replace anti-holomorphic by holomorphic indices.

We will now provide evidence, but not a rigorous proof, for the following claim: If $T_{ijkl}$ purely consists of quantities carrying at least one index, that is no scalar functions appear, then $\mathcal{Z}$ is a constant. If, one the other hand, explicit scalar quantities, such as $K$ or the curvature $R$ appear, in general this no longer holds.

Let us begin by investigating the possible structure of $T_{ijkl}$. Since the superpotential is a constant, we can assume that the coupling tensor is built entirely out of derivatives of $K$. The following list contains the simplest conceivable objects that can be constructed this way: 
\begin{align}
 \label{eq:Zconst_1} & T_{ijkl} = K_{ik} K_{jl} + K_{il} K_{jk} \\
 \label{eq:Zconst_2} & T_{ijkl} = K_i K_j K_k K_l \\
 \label{eq:Zconst_3} & T_{ijkl} = K_i K_k K_{jl} + \text{symmetrized} \\
 \label{eq:Zconst_4} & T_{ijkl} = R_{ijkl} = K_{ijkl} - K_{ijm} K^{mn} K_{nkl} \\
 \label{eq:Zconst_5} & T_{ijkl} = R_{ik} R_{jl} + R_{il} R_{jk} \\
 \label{eq:Zconst_6} & T_{ijkl} = R_{ik} K_{jl} +  \text{symmetrized} \\
 \label{eq:Zconst_7} & T_{ijkl} = R_{ik} K_j K_l +  \text{symmetrized} \\
 \label{eq:Zconst_8} & T_{ijkl} = K_j \nabla_{l} R_{ik} +  \text{symmetrized} \\
 \label{eq:Zconst_9} & T_{ijkl} = \nabla_{j} \nabla_{l} R_{ik} +  \text{symmetrized} 
\end{align}
Here $R_{ijkl}$ denotes the Riemann tensor, $R_{ij}$ the Ricci tensor and $\nabla_k$ the covariant derivative. We will show that for any four-tensor in the upper list of choices  $\mathcal{Z}$ is a constant. For the tensors in eq.~\eqref{eq:Zconst_1} to eq.~\eqref{eq:Zconst_3} this simply follows from the no-scale condition $K^i K_i = 3$. Note that, if we choose $T_{ijkl}$ according to eq.~\eqref{eq:Zconst_4}, then $\mathcal{Z}$ describes the holomorphic sectional curvature along $K^i$. 

The following identity is essential in order to prove our claim
\begin{equation}
\label{eq:appendix_riemann_claim3}
 K_{i_1 \dots i_n j_1 \dots j_m} K^{i_1} \dots K^{i_n} \propto K_{j_1 \dots j_m} \ .
\end{equation}
This relation can be shown stepwise. To begin with note that $\mathcal{V}$ is a homogeneous function of degree $(3/2)$ in the four-cycle volumes $\tau_i$. According to Euler's theorem for homogeneous functions it, thus, has to satisfy
\begin{equation}
 \frac{3}{2} \mathcal{V} = \sum_i \tau_i \mathcal{V}_{i} \ .
\end{equation}
Taking iterative derivatives of this equation we obtain
\begin{equation}
\label{eq:auxiliary_condition_derivatives_proof}
 \sum_i \tau_i \mathcal{V}_{ij_1 \dots j_n} = \frac{3-2n}{2} \mathcal{V}_{j_1 \dots j_n}\ .
\end{equation}
With this we can prove the following auxiliary result\footnote{For $n=2$ this simply corresponds to the no-scale condition.}
\begin{equation}
\label{eq:appendix_riemann_claim1}
 K_{i_1 \dots i_n} K^{i_1} \dots K^{i_n} = \mathrm{const}\ .
\end{equation}
First note that we have
\begin{equation}
\label{eq:Kahler_derivative_upperindex}
 K^i = K^{ij} K_j = - \tau_i\ .
\end{equation}
In general the derivative is of the form
\begin{equation}
 K_{i_1 \dots i_n} = - \frac{2}{\mathcal{V}} \mathcal{V}_{i_1 \dots i_n} + \frac{2}{\mathcal{V}^2} (\mathcal{V}_{i_1 \dots i_{n-1}} \mathcal{V}_{i_n} + \text{symm.}) + \dots + 2\frac{(-1)^{n}(n-1)!}{\mathcal{V}^n} \mathcal{V}_{i_1} \dots \mathcal{V}_{i_n}\ .
\end{equation}
For each term a successive insertion of eq.~\eqref{eq:auxiliary_condition_derivatives_proof} yields precisely the correct power of $\mathcal{V}$, since there are always as many products of derivatives of $\mathcal{V}$ in the numerator as there are powers of $\mathcal{V}$ in the denominator. Thus, one is left with a combinatorial constant for each term. We conclude that eq.~\eqref{eq:appendix_riemann_claim1} is satisfied. 

Now we are in a position to show the following
\begin{equation}
\label{eq:appendix_riemann_claim2}
 K_{i_1 \dots i_n j} K^{i_1} \dots K^{i_n} \propto K_j\ .
\end{equation}
This can be seen via induction in $n$. For $n=1$ the above can simply be checked using eq.~\eqref{eq:auxiliary_condition_derivatives_proof}. Suppose the statement is true for $(n-1)$. Then, taking the derivative of eq.~\eqref{eq:appendix_riemann_claim1} with respect to $\tau_j$, we obtain
\begin{equation}
 K_{i_1 \dots i_n j} K^{i_1} \dots K^{i_n} = - K_{j i_2 \dots i_n} K^{i_2} \dots K^{i_n} -  K_{i_1 j i_3 \dots i_n} K^{i_1} K^{i_3} \dots K^{i_n} - \dots
\end{equation}
Thus, since the statement is true for $(n-1)$, one infers that eq.~\eqref{eq:appendix_riemann_claim2} holds. Now we are in a position to generalize this statement for eq.~\eqref{eq:appendix_riemann_claim3}. Again the proof uses 
induction: For $n=1$ this can be directly deduced by taking derivatives of eq.~\eqref{eq:appendix_riemann_claim2}. For arbitrary $n$ successive differentiation of eq.~\eqref{eq:appendix_riemann_claim2} yields eq.~\eqref{eq:appendix_riemann_claim3}, if eq.~\eqref{eq:appendix_riemann_claim3} holds for $(n-1)$.

Now let us consider for example $\mathcal{Z}$ with $T_{ijkl}$ given by eq.~\eqref{eq:Zconst_4}, then iterative use of eq.~\eqref{eq:appendix_riemann_claim3} yields
\begin{equation}
 \mathcal{Z} \propto K_i K^{ij} K_j + \text{const}.\ ,
\end{equation}
which again gives a constant due to the no-scale property. Similarly one can show that $\mathcal{Z}$ is a constant for the choices in eq.~\eqref{eq:Zconst_5},~\eqref{eq:Zconst_6},~\eqref{eq:Zconst_7}. The cases of eq.~\eqref{eq:Zconst_8} and eq.~\eqref{eq:Zconst_9} require a little more effort, but can be derived making use of properties, such as $(\partial_k K^{ij}) K_{ij} = -K^{ij} K_{ijk}$.
\bibliographystyle{JHEP}
\bibliography{CLW_final_v2}

\providecommand{\href}[2]{#2}\begingroup\raggedright\begin{thebibliography}{10}

\bibitem{Cecotti:1986jy}
S.~Cecotti, S.~Ferrara, and L.~Girardello, {\it {Structure of the Scalar
  Potential in General $N=1$ Higher Derivative Supergravity in
  Four-dimensions}},  {\em Phys.Lett.} {\bf B187} (1987) 321.

\bibitem{Khoury:2010gb}
J.~Khoury, J.-L. Lehners, and B.~Ovrut, {\it {Supersymmetric P(X,$\phi$) and
  the Ghost Condensate}},  {\em Phys.Rev.} {\bf D83} (2011) 125031,
  [\href{http://xxx.lanl.gov/abs/1012.3748}{{\tt arXiv:1012.3748}}].

\bibitem{Koehn:2012ar}
M.~Koehn, J.-L. Lehners, and B.~A. Ovrut, {\it {Higher-Derivative Chiral
  Superfield Actions Coupled to N=1 Supergravity}},  {\em Phys.Rev.} {\bf D86}
  (2012) 085019, [\href{http://xxx.lanl.gov/abs/1207.3798}{{\tt
  arXiv:1207.3798}}].

\bibitem{Farakos:2012qu}
F.~Farakos and A.~Kehagias, {\it {Emerging Potentials in Higher-Derivative
  Gauged Chiral Models Coupled to N=1 Supergravity}},  {\em JHEP} {\bf 1211}
  (2012) 077, [\href{http://xxx.lanl.gov/abs/1207.4767}{{\tt
  arXiv:1207.4767}}].

\bibitem{Farakos:2013zsa}
F.~Farakos, S.~Ferrara, A.~Kehagias, and M.~Porrati, {\it {Supersymmetry
  Breaking by Higher Dimension Operators}},  {\em Nucl.Phys.} {\bf B879} (2014)
  348--369, [\href{http://xxx.lanl.gov/abs/1309.1476}{{\tt arXiv:1309.1476}}].

\bibitem{Sasaki:2012ka}
S.~Sasaki, M.~Yamaguchi, and D.~Yokoyama, {\it {Supersymmetric DBI inflation}},
   {\em Phys.Lett.} {\bf B718} (2012) 1--4,
  [\href{http://xxx.lanl.gov/abs/1205.1353}{{\tt arXiv:1205.1353}}].

\bibitem{Koehn:2012np}
M.~Koehn, J.-L. Lehners, and B.~A. Ovrut, {\it {DBI Inflation in N=1
  Supergravity}},  {\em Phys.Rev.} {\bf D86} (2012) 123510,
  [\href{http://xxx.lanl.gov/abs/1208.0752}{{\tt arXiv:1208.0752}}].

\bibitem{Farakos:2013cqa}
F.~Farakos, A.~Kehagias, and A.~Riotto, {\it {On the Starobinsky Model of
  Inflation from Supergravity}},  {\em Nucl.Phys.} {\bf B876} (2013) 187--200,
  [\href{http://xxx.lanl.gov/abs/1307.1137}{{\tt arXiv:1307.1137}}].

\bibitem{Koehn:2013upa}
M.~Koehn, J.-L. Lehners, and B.~A. Ovrut, {\it {Cosmological super-bounce}},
  {\em Phys.Rev.} {\bf D90} (2014), no.~2 025005,
  [\href{http://xxx.lanl.gov/abs/1310.7577}{{\tt arXiv:1310.7577}}].

\bibitem{Gwyn:2014wna}
R.~Gwyn and J.-L. Lehners, {\it {Non-Canonical Inflation in Supergravity}},
  {\em JHEP} {\bf 1405} (2014) 050,
  [\href{http://xxx.lanl.gov/abs/1402.5120}{{\tt arXiv:1402.5120}}].

\bibitem{Antoniadis:2007xc}
I.~Antoniadis, E.~Dudas, and D.~Ghilencea, {\it {Supersymmetric Models with
  Higher Dimensional Operators}},  {\em JHEP} {\bf 0803} (2008) 045,
  [\href{http://xxx.lanl.gov/abs/0708.0383}{{\tt arXiv:0708.0383}}].

\bibitem{Dudas:2015vka}
E.~Dudas and D.~Ghilencea, {\it {Effective operators in SUSY, superfield
  constraints and searches for a UV completion}},
  \href{http://xxx.lanl.gov/abs/1503.0831}{{\tt arXiv:1503.0831}}.

\bibitem{Buchbinder:1994iw}
I.~Buchbinder, S.~Kuzenko, and Z.~Yarevskaya, {\it {Supersymmetric effective
  potential: Superfield approach}},  {\em Nucl.Phys.} {\bf B411} (1994)
  665--692.

\bibitem{Pickering:1996he}
A.~Pickering and P.~C. West, {\it {The One loop effective superpotential and
  nonholomorphicity}},  {\em Phys.Lett.} {\bf B383} (1996) 54--62,
  [\href{http://xxx.lanl.gov/abs/hep-th/9604147}{{\tt hep-th/9604147}}].

\bibitem{Kuzenko:2014ypa}
S.~M. Kuzenko and S.~J. Tyler, {\it {The one-loop effective potential of the
  Wess-Zumino model revisited}},  {\em JHEP} {\bf 1409} (2014) 135,
  [\href{http://xxx.lanl.gov/abs/1407.5270}{{\tt arXiv:1407.5270}}].

\bibitem{Simon:1990hd}
J.~Z. Simon, {\it Higher-derivative lagrangians, nonlocality, problems, and
  solutions},  {\em Phys. Rev. D} {\bf 41} (Jun, 1990) 3720--3733.

\bibitem{Giddings:2001yu}
S.~B. Giddings, S.~Kachru, and J.~Polchinski, {\it {Hierarchies from fluxes in
  string compactifications}},  {\em Phys.Rev.} {\bf D66} (2002) 106006,
  [\href{http://xxx.lanl.gov/abs/hep-th/0105097}{{\tt hep-th/0105097}}].

\bibitem{Dasgupta:1999ss}
K.~Dasgupta, G.~Rajesh, and S.~Sethi, {\it {M theory, orientifolds and G -
  flux}},  {\em JHEP} {\bf 9908} (1999) 023,
  [\href{http://xxx.lanl.gov/abs/hep-th/9908088}{{\tt hep-th/9908088}}].

\bibitem{Becker:2002nn}
K.~Becker, M.~Becker, M.~Haack, and J.~Louis, {\it {Supersymmetry breaking and
  alpha-prime corrections to flux induced potentials}},  {\em JHEP} {\bf 0206}
  (2002) 060, [\href{http://xxx.lanl.gov/abs/hep-th/0204254}{{\tt
  hep-th/0204254}}].

\bibitem{Kachru:2003aw}
S.~Kachru, R.~Kallosh, A.~D. Linde, and S.~P. Trivedi, {\it {De Sitter vacua in
  string theory}},  {\em Phys.Rev.} {\bf D68} (2003) 046005,
  [\href{http://xxx.lanl.gov/abs/hep-th/0301240}{{\tt hep-th/0301240}}].

\bibitem{Balasubramanian:2005zx}
V.~Balasubramanian, P.~Berglund, J.~P. Conlon, and F.~Quevedo, {\it
  {Systematics of moduli stabilisation in Calabi-Yau flux compactifications}},
  {\em JHEP} {\bf 0503} (2005) 007,
  [\href{http://xxx.lanl.gov/abs/hep-th/0502058}{{\tt hep-th/0502058}}].

\bibitem{Douglas:2006es}
M.~R. Douglas and S.~Kachru, {\it {Flux compactification}},  {\em
  Rev.Mod.Phys.} {\bf 79} (2007) 733--796,
  [\href{http://xxx.lanl.gov/abs/hep-th/0610102}{{\tt hep-th/0610102}}].

\bibitem{Denef:2008wq}
F.~Denef, {\it {Les Houches Lectures on Constructing String Vacua}},
  \href{http://xxx.lanl.gov/abs/0803.1194}{{\tt arXiv:0803.1194}}.

\bibitem{Baumann:2014nda}
D.~Baumann and L.~McAllister, {\it {Inflation and String Theory}},
  \href{http://xxx.lanl.gov/abs/1404.2601}{{\tt arXiv:1404.2601}}.

\bibitem{Antoniadis:1997eg}
I.~Antoniadis, S.~Ferrara, R.~Minasian, and K.~Narain, {\it {R**4 couplings in
  M and type II theories on Calabi-Yau spaces}},  {\em Nucl.Phys.} {\bf B507}
  (1997) 571--588, [\href{http://xxx.lanl.gov/abs/hep-th/9707013}{{\tt
  hep-th/9707013}}].

\bibitem{Antoniadis:2003sw}
I.~Antoniadis, R.~Minasian, S.~Theisen, and P.~Vanhove, {\it {String loop
  corrections to the universal hypermultiplet}},  {\em Class.Quant.Grav.} {\bf
  20} (2003) 5079--5102, [\href{http://xxx.lanl.gov/abs/hep-th/0307268}{{\tt
  hep-th/0307268}}].

\bibitem{Wess:1992cp}
J.~Wess and J.~Bagger, {\em {Supersymmetry and Supergravity}}.
\newblock Princeton Series in Physics. {Princeton Univ. Press}, 1992.

\bibitem{Burgess:2007pt}
C.~Burgess, {\it {Introduction to Effective Field Theory}},  {\em
  Ann.Rev.Nucl.Part.Sci.} {\bf 57} (2007) 329--362,
  [\href{http://xxx.lanl.gov/abs/hep-th/0701053}{{\tt hep-th/0701053}}].

\bibitem{Jaen:1986xl}
X.~Ja\'en, J.~Llosa, and A.~Molina, {\it A reduction of order two for
  infinite-order lagrangians},  {\em Phys. Rev. D} {\bf 34} (Oct, 1986)
  2302--2311.

\bibitem{Nelson:1993nf}
A.~E. Nelson and N.~Seiberg, {\it {R symmetry breaking versus supersymmetry
  breaking}},  {\em Nucl.Phys.} {\bf B416} (1994) 46--62,
  [\href{http://xxx.lanl.gov/abs/hep-ph/9309299}{{\tt hep-ph/9309299}}].

\bibitem{Baumann:2011nm}
D.~Baumann and D.~Green, {\it {Supergravity for Effective Theories}},  {\em
  JHEP} {\bf 1203} (2012) 001, [\href{http://xxx.lanl.gov/abs/1109.0293}{{\tt
  arXiv:1109.0293}}].

\bibitem{Burton:1989ai}
J.~W. Burton, M.~K. Gaillard, and V.~Jain, {\it {Effective one loop scalar
  lagrangian in no scale supergravity models}},  {\em Phys.Rev.} {\bf D41}
  (1990) 3118--3148.

\bibitem{Berg:2005ja}
M.~Berg, M.~Haack, and B.~Kors, {\it {String loop corrections to Kahler
  potentials in orientifolds}},  {\em JHEP} {\bf 0511} (2005) 030,
  [\href{http://xxx.lanl.gov/abs/hep-th/0508043}{{\tt hep-th/0508043}}].

\bibitem{Berg:2007wt}
M.~Berg, M.~Haack, and E.~Pajer, {\it {Jumping Through Loops: On Soft Terms
  from Large Volume Compactifications}},  {\em JHEP} {\bf 0709} (2007) 031,
  [\href{http://xxx.lanl.gov/abs/0704.0737}{{\tt arXiv:0704.0737}}].

\bibitem{Conlon:2005ki}
J.~P. Conlon, F.~Quevedo, and K.~Suruliz, {\it {Large-volume flux
  compactifications: Moduli spectrum and D3/D7 soft supersymmetry breaking}},
  {\em JHEP} {\bf 0508} (2005) 007,
  [\href{http://xxx.lanl.gov/abs/hep-th/0505076}{{\tt hep-th/0505076}}].

\bibitem{Grimm:2013gma}
T.~W. Grimm, R.~Savelli, and M.~Weissenbacher, {\it {On $\alpha'$ corrections
  in N=1 F-theory compactifications}},  {\em Phys.Lett.} {\bf B725} (2013)
  431--436, [\href{http://xxx.lanl.gov/abs/1303.3317}{{\tt arXiv:1303.3317}}].

\bibitem{Grimm:2013bha}
T.~W. Grimm, J.~Keitel, R.~Savelli, and M.~Weissenbacher, {\it {From M-theory
  higher curvature terms to $\alpha'$ corrections in F-theory}},
  \href{http://xxx.lanl.gov/abs/1312.1376}{{\tt arXiv:1312.1376}}.

\bibitem{Junghans:2014zla}
D.~Junghans and G.~Shiu, {\it {Brane curvature corrections to the $ \mathcal{N}
  =$ 1 type II/F-theory effective action}},  {\em JHEP} {\bf 1503} (2015) 107,
  [\href{http://xxx.lanl.gov/abs/1407.0019}{{\tt arXiv:1407.0019}}].

\bibitem{Kehagias:1997cq}
A.~Kehagias and H.~Partouche, {\it {On the exact quartic effective action for
  the type IIB superstring}},  {\em Phys.Lett.} {\bf B422} (1998) 109--116,
  [\href{http://xxx.lanl.gov/abs/hep-th/9710023}{{\tt hep-th/9710023}}].

\bibitem{Policastro:2006vt}
G.~Policastro and D.~Tsimpis, {\it {R**4, purified}},  {\em Class.Quant.Grav.}
  {\bf 23} (2006) 4753--4780,
  [\href{http://xxx.lanl.gov/abs/hep-th/0603165}{{\tt hep-th/0603165}}].

\bibitem{Liu:2013dna}
J.~T. Liu and R.~Minasian, {\it {Higher-derivative couplings in string theory:
  dualities and the $B$-field}},  {\em Nucl.Phys.} {\bf B874} (2013) 413--470,
  [\href{http://xxx.lanl.gov/abs/1304.3137}{{\tt arXiv:1304.3137}}].

\bibitem{Green:1997as}
M.~B. Green, M.~Gutperle, and P.~Vanhove, {\it {One loop in
  eleven-dimensions}},  {\em Phys.Lett.} {\bf B409} (1997) 177--184,
  [\href{http://xxx.lanl.gov/abs/hep-th/9706175}{{\tt hep-th/9706175}}].

\bibitem{Balasubramanian:2004uy}
V.~Balasubramanian and P.~Berglund, {\it {Stringy corrections to Kahler
  potentials, SUSY breaking, and the cosmological constant problem}},  {\em
  JHEP} {\bf 0411} (2004) 085,
  [\href{http://xxx.lanl.gov/abs/hep-th/0408054}{{\tt hep-th/0408054}}].

\bibitem{Cicoli:2007xp}
M.~Cicoli, J.~P. Conlon, and F.~Quevedo, {\it {Systematics of String Loop
  Corrections in Type IIB Calabi-Yau Flux Compactifications}},  {\em JHEP} {\bf
  0801} (2008) 052, [\href{http://xxx.lanl.gov/abs/0708.1873}{{\tt
  arXiv:0708.1873}}].

\bibitem{Louis:2012nb}
J.~Louis, M.~Rummel, R.~Valandro, and A.~Westphal, {\it {Building an explicit
  de Sitter}},  {\em JHEP} {\bf 1210} (2012) 163,
  [\href{http://xxx.lanl.gov/abs/1208.3208}{{\tt arXiv:1208.3208}}].

\bibitem{Berg:2005yu}
M.~Berg, M.~Haack, and B.~Kors, {\it {On volume stabilization by quantum
  corrections}},  {\em Phys.Rev.Lett.} {\bf 96} (2006) 021601,
  [\href{http://xxx.lanl.gov/abs/hep-th/0508171}{{\tt hep-th/0508171}}].

\bibitem{Cicoli:2008va}
M.~Cicoli, J.~P. Conlon, and F.~Quevedo, {\it {General Analysis of LARGE Volume
  Scenarios with String Loop Moduli Stabilisation}},  {\em JHEP} {\bf 0810}
  (2008) 105, [\href{http://xxx.lanl.gov/abs/0805.1029}{{\tt
  arXiv:0805.1029}}].

\bibitem{Candelas:1990pi}
P.~Candelas and X.~de~la Ossa, {\it {Moduli Space of {Calabi-Yau} Manifolds}},
  {\em Nucl.Phys.} {\bf B355} (1991) 455--481.

\bibitem{Cicoli:2013swa}
M.~Cicoli, J.~P. Conlon, A.~Maharana, and F.~Quevedo, {\it {A Note on the
  Magnitude of the Flux Superpotential}},  {\em JHEP} {\bf 1401} (2014) 027,
  [\href{http://xxx.lanl.gov/abs/1310.6694}{{\tt arXiv:1310.6694}}].

\bibitem{Candelas:1994hw}
P.~Candelas, A.~Font, S.~H. Katz, and D.~R. Morrison, {\it {Mirror symmetry for
  two parameter models. 2.}},  {\em Nucl.Phys.} {\bf B429} (1994) 626--674,
  [\href{http://xxx.lanl.gov/abs/hep-th/9403187}{{\tt hep-th/9403187}}].

\bibitem{Denef:2004dm}
F.~Denef, M.~R. Douglas, and B.~Florea, {\it {Building a better racetrack}},
  {\em JHEP} {\bf 0406} (2004) 034,
  [\href{http://xxx.lanl.gov/abs/hep-th/0404257}{{\tt hep-th/0404257}}].

\bibitem{Grimm:2014efa}
T.~W. Grimm, T.~G. Pugh, and M.~Weissenbacher, {\it {The effective action of
  warped M-theory reductions with higher derivative terms - Part I}},
  \href{http://xxx.lanl.gov/abs/1412.5073}{{\tt arXiv:1412.5073}}.

\bibitem{Martucci:2014ska}
L.~Martucci, {\it {Warping the K\"{a}hler potential of F-theory/IIB flux
  compactifications}},  {\em JHEP} {\bf 1503} (2015) 067,
  [\href{http://xxx.lanl.gov/abs/1411.2623}{{\tt arXiv:1411.2623}}].

\bibitem{Fulling:1992vm}
S.~Fulling, R.~C. King, B.~Wybourne, and C.~Cummins, {\it {Normal forms for
  tensor polynomials. 1: The Riemann tensor}},  {\em Class.Quant.Grav.} {\bf 9}
  (1992) 1151--1197.

\bibitem{Freeman:1986}
M.~Freeman and C.~Pope, {\it Beta-functions and superstring compactifications},
   {\em Physics Letters B} {\bf 174} (1986) 48--50.

\bibitem{CHEN01011975}
B.-Y. Chen and K.~Ogiue, {\it Some characterizations of complex space forms in
  terms of chern classes},  {\em The Quarterly Journal of Mathematics} {\bf 26}
  (1975) 459--464.

\bibitem{kobayashi1969foundations}
S.~Kobayashi and K.~Nomizu, {\em Foundations of Differential Geometry: Vol.:
  2}.
\newblock Interscience Tracts in Pure and Applied Mathematics. Interscience
  Publishers, 1969.

\bibitem{Covi:2008ea}
L.~Covi, M.~Gomez-Reino, C.~Gross, J.~Louis, G.~A. Palma, et~al., {\it {de
  Sitter vacua in no-scale supergravities and Calabi-Yau string models}},  {\em
  JHEP} {\bf 0806} (2008) 057, [\href{http://xxx.lanl.gov/abs/0804.1073}{{\tt
  arXiv:0804.1073}}].

\end{thebibliography}\endgroup

\end{document}